\documentclass[9pt,twocolumn,twoside]{osajnl}

\journal{jocn} 

\setboolean{shortarticle}{false}
\title{Real-Time Capable, Low-latency Upstream Scheduling in Multi-Tenant, SLA Compliant TWDM PON}

\author[1, *]{Arijeet Ganguli}
\author[1]{Marco Ruffini}

\affil[1]{School of Computer Science and Statistics, Trinity College Dublin, Ireland}

\affil[*]{gangulia@tcd.ie}

\usepackage{float}
\usepackage{graphicx}

\usepackage{subcaption} % For subfigures
\usepackage{cleveref}
\usepackage{nameref}

\usepackage{algorithm}% http://ctan.org/pkg/algorithms
\usepackage{algpseudocode}% http://ctan.org/pkg/algorithmicx

\usepackage{legacy-styles/jocn}

\graphicspath{{./figures/}}

\begin{abstract}
Virtualized Passive Optical Networks (vPONs) offer a promising solution for modern access networks, bringing enhanced flexibility, reduced capital expenditures (CapEx), and support for multi-tenancy. By decoupling network functions from physical infrastructure, vPONs enable service providers to efficiently share network resources among multiple tenants. In this paper, we propose a novel merging DBA algorithm, called the Dynamic Time and Wavelength Allocation (DTWA) algorithm, for a virtualized DBA (vDBA) architecture in multi-tenant PON environments. The Algorithm, which enables the merging of multiple virtual DBAs into a physical bandwidth map, introduces multi-channel support, allowing each Optical Network Unit (ONU) to dynamically change, taking into consideration different switching times, transmission wavelength. Leveraging the Numba APIs for high-performance optimization, the algorithm achieves real-time performance with minimal additional latency, meeting the stringent requirements of SLA-compliant, latency-critical 6G applications and services. Our analysis highlights an important trade-off in terms of throughput in multi-tenant conditions, between single-channel vs. multi-channel PONs, as a function of ONUs tuning time. We also compare the performance of our algorithm for different traffic distributions. 
Finally, in order to assess the time computing penalty of dynamic wavelength optimisation in the merging DBA algorithm, we compare it against a baseline Static Wavelength Allocation (SWA) algorithm, where ONUs are designated a fixed wavelength for transmission. %Our results show that while SWA performs similar to DTWA when the tuning time is way higher than the average size of the allocation. 
\end{abstract}

\setboolean{displaycopyright}{false} % Do not include copyright or licensing information in submission.

\begin{document}

\maketitle

\section{Introduction}
\label{introduction}
As the development of 6G networks takes pace, the telecommunication industry faces numerous challenges to accommodate exponential data traffic, massive connectivity, and increasingly diverse quality of service (QoS) requirements. 6G builds on the foundation of 5G while pushing the boundaries of network capabilities to support ultra-high throughput, extremely low end-to-end latency (of the order of $1 ms$) \cite{5G_survey}, and deterministic QoS for a new era of services and applications. Passive Optical Networks (PONs) have emerged as a critical technology in the evolution of these access networks, offering a cost-effective, scalable, reliable, and high-speed solution for delivering broadband services to a wide range of users \cite{PON_resiliency}. With the demand for higher bandwidth and lower latency, particularly in the context of 5G and beyond, PONs are playing an increasingly vital role in both residential and enterprise applications. 
Their ability to deliver symmetric data rates, such as in the multi-channel 40G and single channel 50G-PON \cite{ITU40GPON, ITU50GPON}, makes them a suitable candidate for supporting next-generation applications, including 6G fronthaul and other latency-sensitive services.

Traditional PON architectures are somewhat rigid, with a single network operator controlling the optical line terminal (OLT) and managing bandwidth allocation through a centralized Dynamic Bandwidth Allocation (DBA) mechanism. But over the past decade, network architectures have undergone significant transition from closed systems to open and disaggregated systems. Software Defined Networking (SDN) and Network Function Virtualization (NFV) have played a key role, offering programmability, cost efficiency and flexibility. PONs have experienced a similar transition towards virtualization \cite{mec,fasa_ntt}, which enables dynamic software-based control of scheduling algorithms and multi-tenancy. This allows different Virtual Network Operators (VNOs) to run multiple independent upstream scheduling algorithms in a shared PON. This maximizes the usage of available bandwidth while significantly reducing operational costs through sharing of network resources, making it a highly scalable and flexible solution for next-generation network services \cite{multi_tenancy}.  

Due to their cost-effectiveness and high scalability, vPONs can be used to support mobile fronthaul technology for Open Radio Access Network (O-RAN) \cite{o_ran_fronthaul}. This can be achieved by connecting small cells in 5G functional split architecture (i.e., supporting ORAN 7.2 Remote Unit(RU)-Distributed Unit(DU) split as well as higher level splits) using a PON point-to-multipoint (P2MP) topology. This can help to reduce the Capital Expenditure (CAPEX) and Operational Expenditure (OPEX) for the fronthaul network compared to the traditional point-to-point (P2P) topology. However, P2MP fronthaul introduces additional latency within the PON system due to upstream scheduling, as typically the PON and DU upstream schedulers operate independently. So, in order to support applications that require Ultra-Reliable Low-Latency Communication (URLLC), the concept of Cooperative DBA was originally introduced in \cite{cooperative_dba}. This focuses on the development of an optical-mobile cooperative interface that translates mobile upstream scheduling data into PON transmit request information. The concept was later standardised as Cooperative Transport Interface (CTI) \cite{cti}, which facilitates enhanced coordination between PON and RAN upstream schedulers, leading to more efficient resource allocation and reduced latency. %It is also necessary that the scheduling mechanism implement real-time DBA algorithms (in range of few ten microseconds) \cite{5G_network_slicing}.
%supported by the cooperative DBA \cite{cooperative_dba}, using the Cooperative Transport Interface (CTI) \cite{cti}, which introduces additional network latency. Hence, to support applications that require Ultra-Reliable Low-Latency Communication (URLLC), it is necessary that the mechanism supports real-time scheduling algorithms (in range of few ten microseconds) \cite{5G_network_slicing}.

\subsection{Literature Review}
Early DBA algorithms in traditional PON architectures have focused primarily on fairness and efficiency. The most common approach has been the Grant-Request model, where each ONU sends a bandwidth request to the Optical Line Terminal (OLT), which then allocates the available bandwidth. The classical Interleaved Polling with Adaptive Cycle Time (IPACT) algorithm, introduced in \cite{kramer2002ipact}, is one of the foundational DBA schemes used in PONs. IPACT dynamically adjusts polling cycles based on traffic conditions, reducing idle time and improving bandwidth utilization. However, while IPACT provides an efficient framework for bandwidth allocation, it faces challenges in scenarios with diverse traffic patterns, especially when serving multiple services with distinct QoS requirements.

Subsequent DBA algorithms introduced hierarchical bandwidth allocation schemes, based on traffic priorities. One of such network traffic prioritization algorithm is presented in \cite{giant_dba} where optimized versions of GigaPON access network (GIANT) DBA called deficit XGIANT (XGIANT-D) and proportional XGIANT (XGIANT-P) were proposed. It was demonstrated that both XGIANT-D and XGIANT-P ensure strictly prioritized, low, and fair mean queuing delays for aggregated voice, video, and best-effort upstream traffic in two loaded conditions in 10-Gbit-capable PONs (XG-PONs). Moreover, XGIANT-D and XGIANT-P also ensure reduced packet losses for aggregated upstream traffic in the XG-PON-based LTE backhaul compared to GIANT and efficient bandwidth utilization algorithms. Another novel DBA algorithm supporting QoS and a services hierarchy is proposed in \cite{dba_yang} and divides network services according to different priorities, resulting in improved channel utilization and reduced average delay. 

As PONs evolved to accommodate multi-service networks, research work on DBA algorithms has started to incorporate Service Level Agreement (SLA)-based bandwidth allocation mechanisms to ensure compliance with contracted QoS levels. SLA-aware DBA algorithms consider the bandwidth requests from ONUs and prioritize traffic based on latency and bandwidth guarantees, ensuring higher-priority traffic, such as video streaming or 5G fronthaul, receives timely allocation. For instance, the Delay-Aware DBA (DA-DBA) algorithm \cite{da_dba} enhances traditional DBA schemes by factoring in the delay sensitivity of different traffic types. The DA-DBA approach dynamically adjusts bandwidth allocations based on the delay profiles of active flows, ensuring that latency-sensitive traffic is prioritized, while still optimizing overall network efficiency. Another important development is the Predictive DBA, which leverages traffic forecasting to allocate bandwidth in advance based on predicted demand \cite{predictive_dba}. 

The next evolutionary step was enabling multi-tenancy by allowing multiple service providers to share infrastructure. True multi-tenancy necessitates full virtualization of the PON network, allowing VNOs to control the PON protocol and perform capacity scheduling as if they owned and controlled the physical OLT devices \cite{multi_tenant_pon}. In this scenario, service providers have more control over their DBAs, which are designed and customized to support a broader range of service agreements, such as low latency. 
In this regards, authors in \cite{pon_frame_li} introduced an inter-frame bandwidth resource-sharing framework for XG-PONs, where a fixed time slot frame of 125 µs is considered a slice. An operator is assigned an entire slice based on the proposed algorithm. This proposed framework allows for greater customizability, network isolation, and efficient bandwidth utilization. Nonetheless, it is observed that this solution is susceptible to greater delays, since bandwidth is likely to be wasted due to underutilized frames by specific operators. The authors extend this work in \cite{li_second} to schedule whole frames to individual VNOs in a round-robin fashion. This approach, however, lacks the ability to mix multi-vendor allocations within each frame, resulting in a latency that is dependent on the number of VNOs sharing the infrastructure. In \cite{merging_last_jocn}, the authors present a novel dynamic merging algorithm that merge multiple bwmaps based on traffic priority like Assured, Best Effort, Unassured etc. In \cite{twdm_pon_last}, the authors present a novel load adaptive merging algorithm (LAMA) that modifies the existing strict priority scheme called the Priority-Based Merging Algorithm (PBMA) scheme and improves the performance of the merging engine by allocating the in a load adaptive manner to the in the multi-tenant PON architecture. 

All the work mentioned above considers strict time invariant traffic prioritization approach while allocating DBA.  Different packets are marked with different priority levels and then queue management algorithms are applied to determine how packet prioritization is applied. While these can be effective in assigning relative importance to different services, they are unable to capture the complexity of specific SLAs, especially in a heterogeneous and shared network environment. This shortcoming leads to inefficient utilization of resources and inability to guarantee performance. In \cite{stateful_sla_first}, the authors presented a stateful SLA compliant PON Hypervisor that runs a merging DBA algorithm with a stateful mechanism where the bandwidth allocations are made minimizing the breach of SLAs over time. In \cite{stateful_sla_second}, we present a real-time stateful SLA-compliant merging algorithm capable to run multiple SLA compliant services and applications. 
In \cite{dynamic_twdm}, we present a multi-tenant multi-wavelength upstream transmission scheme for virtualised PONs, enabling compliance with latency-oriented SLAs called the Dynamic Time and Wavelength Allocation (DTWA) algorithm. The study demonstrated an important trade-off between the latency of multi-channel systems and the transmitters' tuning time. If the tuning time is negligible compared to the burst overhead, the multi-channel system has better latency performance. This is due to the fact that the burst overhead tends to have similar time duration independently of the line rate \cite{ITU50GPON}, thus higher line rate channels require a proportionally larger amount of bits than lower line rate channels. But as the ONU tuning time increases, the single channel system outperforms the multi-channel ones. However, through network traffic simulations, we observed that the dynamic wavelength and time slot allocation algorithm is computationally intensive and added notable additional latency to the PON upstream scheduling mechanism.                   

\subsection{Contribution}
The work presented here is an extension of our previous work (\cite{dynamic_twdm}). Here, we first extend the vDBA architecture for a multi-channel PON network, where the merging of virtual Bandwidth Maps (vBmaps) is carried out across multiple wavelength channels. The key objective is to define a number of SLAs, expressed in terms of maximum latency and compliance level, and to propose an algorithm (realtime DTWA) that can operate the Bandwidth Map merging operation, minimizing the probability to breach SLAs, across all VNOs. The evaluation of our work is based on a few key performance analysis. First, we compare the latency for systems using different line rates and number of channels. Assuming an overall PON capacity of 200Gb/s, we consider a system with 8 channels at 25Gb/s, one with 4 channels running at 50Gb/s, and one with a single channel at 200Gb/s (being single channel there is no multi-wavelength allocation for the 200Gb/s option). Then we show how the difference in performance changes when we consider different tuning times for the ONU transmitters. For this we compare the case of negligible tuning time (i.e., if the system implements channel bonding, \cite{channel_bonding}, so that the ONU has more than one transceiver always ready to transmit at least at another wavelength); the case of Class 1 transmitters (i.e., tuning time $<10 \mu s$) \cite{ITU40GPON}, with tuning time of 250 ns (i.e., close to the burst overhead time) and 1 us; and a Class 2 transmitters (i.e., tuning time $>10 \mu s$ but $<25 ms$), with tuning time of 15 us. Class 3 devices are not considered as they have tuning times longer than 25 ms, which is more than two order of magnitudes higher than the PON frame, and would in practice represent a semi-static allocation.

Then, we extend the results by varying the network traffic distribution from the uniform traffic distribution considered in \cite{dynamic_twdm} to different distributions namely Poisson, Zipf-Mendelbrot and Pareto distribution. We show that the performance of our algorithm have some dependency on the traffic distribution. The performance is higher for uniformly distributed traffic (low variance) with higher expected arrival times. 
We also compare the DTWA algorithm with a Static Wavelength Allocation (SWA) algorithm. On the one hand we assess the improvement in latency that dynamic wavelength brings with respect to a static allocation. On the other hand we also compare the merging engine algorithm execution time for the two algorithms, highlighting an important trade-off. %We use this comparison also to highlight the difference in computing time for the two% advantage brought by trade-off between . In this algorithm, the wavelength allocation is done statically, meaning the ONUs are designated a fixed wavelength for transmission. %The performance analysis of this algorithm showed that the performance is equivalent to the performance of the DTWA algorithm with tuning times higher than the average transmission time of the bmap allocations. 
This work also includes optimisation of runtime execution by leveraging the Just-In-Time (JIT) compiler for Python, Numba, reporting the results for both static and dynamic wavelength allocation algorithms, showing yet another important trade-off between static and dynamic wavelength allocation algorithms.   

\section{System Architecture and proposed approach}
\label{section2}
\begin{figure}[htbp]
\centering
\includegraphics[width=.8\linewidth]{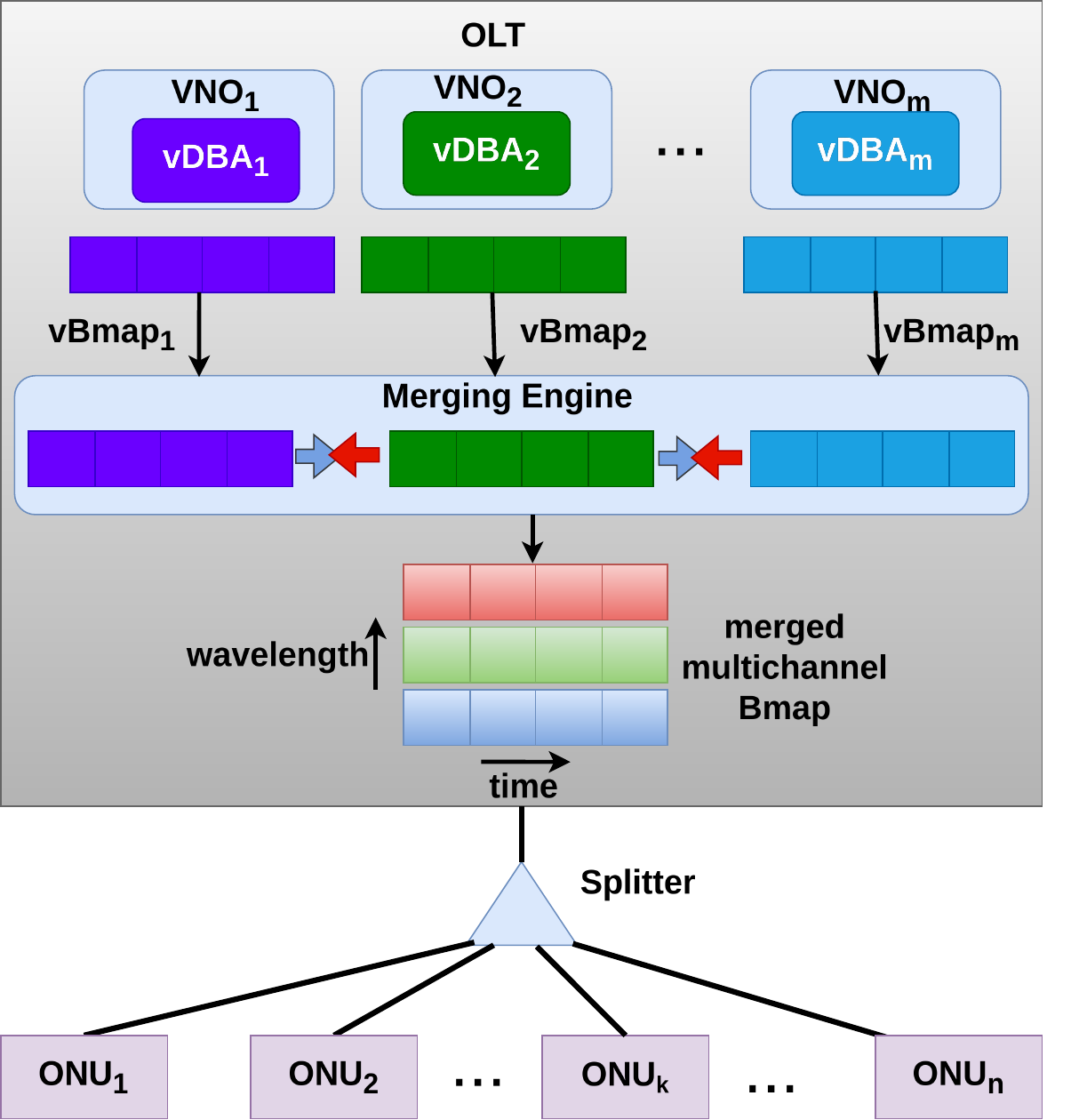}
\caption{vDBA architecture for the multi tenant TWDM PON}
\label{fig:vDBA_PON}
\end{figure}

The virtual DBA architecture for a multi-channel system addressed in this work is shown in Fig. \ref{fig:vDBA_PON}. In our approach, multiple VNOs run different schedulers in parallel (vDBAs), each forwarding a vBMap, which allocates upstream transmission channel and slots to a group of ONUs. 
The scheduling hypervisor (or merging engine) collects all such virtual bandwidth maps (vBmaps) that contain information about the requested bandwidth from different ONUs belonging to the respective VNOs for a specific interval of time. In this multi-wavelength approach, the OLT also has the option to select transmission over different channels, to minimize grant delay, although this is constrained by the ONU tuning time. The merging engine allocates a transmission channel and time slot within the channel for the allocations of all the BMaps. Since, each ONU is tuned to a specific wavelength at any point in time, the OLT can broadcast the allocated channel and time slot information downstream to all the ONUs using vBmaps through each of the transmission channels and the ONUs that listen to that channel can access the vBmap information and can schedule its transmission accordingly. As shown in Fig. \ref{fig:vDBA_PON}, the TWDM system allows transmission in 3 different channels and the merging engine broadcasts the allocated channel and slot information to the respective ONUs in form of vBmaps through each of these channels. It should be noted that downstream and upstream channels can be selected independently for each ONU. Each bandwidth map will contain information on what time slot and upstream wavelength should be used by each ONU. It should also be noted that in this work we do not address how the downstream wavelength is selected, as this can be typically organised by the management layer and is not part of the upstream allocation problem.
Our merging engine resolves allocation conflicts among the vBmaps based on specific Service Level Agreements (SLAs), so that it minimizes the probability of breaching SLAs. The use of a stateful algorithm, which takes into consideration the history of a service flow when making scheduling prioritization decisions, is preferred to stateless algorithms \cite{stateful_sla_first}. This is because a stateful algorithm can prioritize flows depending on how close they are to breaching their specific SLA target \cite{stateful_sla_second}. 
% In this multi-wavelength approach, the OLT also has the option to select transmission over different channels, to minimize grant delay, although this is constrained by the ONU tuning time. Our merging engine resolves allocation conflicts among the vBmaps based on specific Service Level Agreements (SLAs), so that it minimizes the probability of breaching SLAs. The use of a stateful algorithm, which takes into consideration the history of a service flow when making scheduling prioritization decisions, is preferred to stateless algorithms \cite{stateful_sla_first}. This is because a stateful algorithm can prioritize flows depending on how close they are to breaching their specific SLA target \cite{stateful_sla_second}.   

Thus in this work, we propose a heuristic stateful TWDM scheduling algorithm. The objective of the algorithm is to maximize the SLA compliance across all the flows during upstream transmission. We focus specifically on the additional latency introduced by the multi-tenancy sharing aspect of the PON. An SLA breach occurs when a given flow accumulates a number of delayed upstream slots that is above its target SLA threshold. For example for an SLA with maximum merging delay of 25 $\mu s$ with 99\% compliance, every time an upstream slot is delayed by more than 25 $\mu s$ with respect to the requested time slot in the virtual BMap, we increment a counter. If the counter goes above the non-compliance rate (in this case 100-99=1\%), calculated over a number of frames (i.e., we use a 1ms window, which is the time duration of a 5G sub-frame), then we consider that an SLA breach has occurred. 

The problem of upstream wavelength and time slot allocation is formulated as a Mixed Integer Programming (MIP) problem. Equation \ref{eq:maxtime} calculates the maximum delay of any given allocation to remain within the target threshold for SLA breach. We maintain a 4 dimensional binary decision variable matrix \textbf{\textit{X}} where the objective function (equation \ref{eq:objective}) is explained as follows - the inner truth value function calculates the packet level breach and the outer truth value function calculates the flow level breach across each virtual BMap. Equation \ref{eq:constraint1} is the constraint that any particular channel at any time slot transmits at most 1 BMap allocation. Equation \ref{eq:constraint2} is the constraint that all BMap allocations are alloted unique channel time-slot pairs. Equation \ref{eq:constraint3} checks the conservation of BMap allocations within each virtual BMap.
\begin{table}[H]
\centering
\caption{\bf Mathematical Notations}
\begin{tabular}{p{0.8cm} p{7.5cm}} % Adjust the column widths as needed
\hline
Symbol & Description \\
\hline
    $N$ & Number of bmaps from the respective VNOs \\
    $M$ & Number of SLAs \\
    $W$ & Number of channels \\
    $F$ & Frame size in allocation units \\
    $\text{SLA}_{i}$ & SLA ID for the $i^{\text{th}}$ VNO \\
    $\text{lat}_i$ & Maximum allowed packet level latency for the allocations from the $i^{\text{th}}$ VNO \\
    $\text{br}^i_{pt}$ & Maximum fraction of packets in a flow that can breach packet level SLA \\
    $|S_i|$ & Number of allocations in the $i^{\text{th}}$ bmap \\
    $bw_{i,j}$ & $j^{\text{th}}$ bandwidth allocation report from the $i^{\text{th}}$ bmap \\
    $req_{i,j}$ & Start time of $bw_{i,j}$ \\
    $t^{\max}_{i,j}$ & Maximum allowed time $bw_{i,j}$ can be delayed without breaching the packet level SLA \\
    $\mathbb{X}_{i,j,k,l}$ & Binary Decision Variable, $bw_{i,j}$ allotted time slot $l$ of channel $k$ if 1, else 0 \\
    $\mathbb{I}$ & Boolean truth value function \\
\hline
\end{tabular}
\label{tab:notations}
\end{table}

\setlength{\abovedisplayskip}{0pt}  
\setlength{\belowdisplayskip}{0pt}  

\begin{equation}
t_{i,j}^{max} = t_{i,j}^{req} + lat_i \label{eq:maxtime}
\end{equation}
\begin{equation}
\min \sum_{i=1}^{N} \mathbb{I}\left( \frac{1}{|S_i|} \sum_{j=1}^{F} \sum_{k=1}^{W} \sum_{l=1}^{F} \mathbb{X}_{i,j,k,l} \mathbb{I}(l > t_{i,j}^{max}) > br_i^{pt} \right) \label{eq:objective}
\end{equation}
\begin{equation}
\textrm{s.t.} \quad \sum_{i=1}^{N} \sum_{j=1}^{F} \mathbb{X}_{i,j,k,l} \leq 1, \quad \forall \, 1 \leq k \leq W, \, 1 \leq l \leq F \label{eq:constraint1}
\end{equation}
\begin{equation}
\sum_{k=1}^{W} \sum_{l=1}^{F} \mathbb{X}_{i,j,k,l} \doteq 1, \quad \forall \, 1 \leq i \leq N, \, 1 \leq j \leq F \label{eq:constraint2}
\end{equation}
\begin{equation}
\sum_{j=1}^{F} \sum_{k=1}^{W} \sum_{l=1}^{F} \mathbb{X}_{i,j,k,l} \doteq |S_i|, \quad \forall \, 1 \leq i \leq N \label{eq:constraint3}
\end{equation}

In our work, we first propose a heuristic Dynamic Time and Wavelength Allocation (DTWA) algorithm (\ref{alg:dtwa}~\nameref{alg:dtwa}) to allocate wavelength and time slots for upstream transmission in multi-tenant PONs. The dynamic algorithm maintains a few key data structures. A flow SLA breach table keeps track of how much each traffic flow has breached its SLA in percentage (i.e., going above non-compliance rate); this is important to minimize SLA breach, as the algorithm will prioritize scheduling for the flows that have a higher breach of their SLA. A channel freetime table maintains and updates the earliest free time of each channel over time, i.e., when the channel can be reallocated to a different BMap allocation. We also keep track of the current channel tuned on each transceivers in the ONUs (we consider only 1 transceiver per ONU). With reference to the pseudocode \ref{alg:dtwa}, the main procedure \textbf{DynamicTwdm} (lines 38-44) is explained as follows. Everytime it receives the input BMaps from the various VNOs, it first calculates the allocation maxtime using procedure \textbf{CalculateMaxtTime (lines 4-7)} in line 41. The allocation maxtime is the latest time an allocation can be scheduled within its latency target (equation \ref{eq:maxtime}). Then, it merges the bmaps from the VNOs and sorts them based on the following comparison, first, in descending order of bmap allocation flow SLA breach, next, in ascending order of the allocatipn maxtimes and finally ascending order of allocation sizes (line 42). This is explained by the procedure \textbf{SortBmaps} (lines 8-19). Then, the procedure \textbf{AssignResource} (lines 25-34) is called which allocates wavelength and time-slot within that wavelength to each of the sorted bmap allocations (line 43). In the \textbf{AssignResource} procedure, we first find the earliest available wavelength using the sub-procedure \textbf{FindMinIndex} (lines 20-24). Then, we check if no channel has been allocated to the ONU or switching channels is faster than waiting for the current tuned channel to be free for transmission of the next bmap allocation. If this is the case, we allocate the earliest free channel for transmission, else we stay tuned on the same channel for further transmission (lines 29-32). Then, we update the channel freetime table and OnuTunedChannelMap after the channel assignment to the allocation (lines 33) and finally return the corresponding channelwise bmaps for transmission accordingly (line 34). Finally, we update the SLA breach table using the procedure \textbf{UpdateSlaBreachTable} (lines 35-37) in line 44. In this procedure, we calculate the scheduled time for each of the bmap allocations and update the SLA breach of each VNO flow based on how many allocations of the flow are scheduled beyond the respective allocation maxtimes.
\begin{algorithm}[htbp]
\caption{Dynamic Time and Wavelength Allocation Algorithm}
\label{alg:dtwa}
\begin{algorithmic}[1]
\State \textbf{Inputs:} bmaps, nChannels, channelTuningTime, slaTable
\State \textbf{Outputs:} mergedBmap
\State \textbf{State Tables:} slaBreachTable, channelFreeTimeTable, onuFreeTimeTable, onuTunedChannelMap

\Procedure{CalculateMaxTime}{bmaps}
    \For{bmap \textbf{in} bmaps}
        \For{bmapElem \textbf{in} bmap}
            \State bmapElem.maxTime $\gets$ bmapElem.startTime + slaTable[bmapElem.vnoId].maxLatency
        \EndFor
    \EndFor
\EndProcedure

\Procedure{SortBmaps}{bmaps, slaBreachTable}
    \State bmapsMerge $\gets$ Flatten(bmaps)
    \Function{Compare}{elem1, elem2}
        \State slaBreach1 $\gets$ slaBreachTable[elem1.vnoId]
        \State slaBreach2 $\gets$ slaBreachTable[elem2.vnoId]
        \If{slaBreach1 $\neq$ slaBreach2}
            \State \Return slaBreach1 $<$ slaBreach2
        \ElsIf{elem1.maxTime $\neq$ elem2.maxTime}
            \State \Return elem1.maxTime $<$ elem2.maxTime
        \Else
            \State \Return elem1.size $<$ elem2.size
        \EndIf
    \EndFunction
    \State \Return \Call{MergeSort}{bmapsMerge, Compare}
\EndProcedure

\Procedure{FindMinIndex}{arr, A}
    \State MinVal $\gets$ \Call{Min}{arr}
    \State MinIndices $\gets$ Indices of MinVal in arr
    \State BestIndex $\gets$ \Call{ArgMin}{A[MinIndices]}
    \State \Return BestIndex
\EndProcedure

\Procedure{AssignResource}{bmapsSorted, channelFreeTimeTable, onuFreeTimeTable, onuTunedChannelMap, nChannels, channelTuningTime}
    \State Initialize ChannelAllocCtr and BmapsChannel
    \For{elem \textbf{in} bmapsSorted}
        \State Find EarliestFreeChannel using \Call{FindMinIndex}{ChannelFreeTimeTable, ChannelAllocCtr}
        \State SchedTimeSameChannel $\gets$ channelFreeTimeTable[current channel]
        \State SchedTimeDiffChannel $\gets$ \Call{Max}{EarliestFreeChannel, (channelTuningTime + channelFreeTimeTable[current channel])}
        \State \textit{switching\_faster} $\leftarrow$ \textbf{True} \textbf{if} SchedTimeDiffChannel $<$ SchedTimeSameChannel \textbf{else} \textbf{False}  
        \If{elem's current channel unavailable or \textit{switching\_faster}}
            \State Assign EarliestFreeChannel
        \Else
            \State Assign current channel
        \EndIf
        \State Update onuTunedChannelMap, ChannelFreeTimes, and ChannelAllocCtr
    \EndFor
    \State \Return BmapsChannel
\EndProcedure

\Procedure{UpdateSlaBreachTable}{BmapsChannel, slaBreachTable, numAllocsVno}
    \For{bmapChannel \textbf{in} BmapsChannel}
        \State Update schedTime and SLA breaches for each element
    \EndFor
\EndProcedure

\Procedure{DynamicTWDM}{bmaps, slaTable, nChannels, channelTuningTime}
    \State Initialize state tables
    \While{\textbf{true}}
        \State \Call{CalculateMaxTime}{bmaps}
        \State BmapsSorted $\gets$ \Call{SortBmaps}{bmaps, slaBreachTable}
        \State BmapsChannel $\gets$ \Call{AssignResource}{BmapsSorted, ChannelFreeTimeTable, OnuTunedChannelMap, nChannels, channelTuningTime}
        \State \Call{UpdateSlaBreachTable}{BmapsChannel, SlaBreachTable, numAllocsVno}
    \EndWhile
\EndProcedure
\end{algorithmic}
\end{algorithm}

We also present another version, called the Static Wavelength Allocation (SWA) algorithm (\ref{alg:swa}~\nameref{alg:swa}), where we restrict channel switching in ONUs such that once a channel is allocated to an ONU, the ONU keeps transmitting in the same channel. This can be considered as the case with channel tuning time higher than the PON frame duration (i.e., slower Class 2 device and Class 3 devices).
With reference to the pseudocode (algorithm~\ref{alg:swa}), like in the DTWA algorithm, we first calculate the allocation maxtime using procedure \textbf{CalculateMaxtTime (lines 4-7)} in line 33. Then we call the procedure \textbf{SortBmaps} (lines 8-26) in line 34. In this procedure, we first separate the bmap allocations according to different channels they are meant to be transmitted (lines 10-14). Then we sort the list of bmap allocations for the separate channels based on the same comparator function as in the DTWA algorithm (lines 15-25). Finally, we update the SLA breach table in line 35. 
\begin{algorithm}[htbp]
\caption{Static Wavelength Upstream Scheduling Algorithm}
\label{alg:swa}
\begin{algorithmic}[1]
\State \textbf{Inputs:} bmaps, nChannels, slaTable
\State \textbf{Outputs:} mergedBmap
\State \textbf{State Tables:} slaBreachTable, onuTunedChannelMap

\Procedure{CalculateMaxTime}{bmaps}
    \For{bmap \textbf{in} bmaps}
        \For{bmapElem \textbf{in} bmap}
            \State bmapElem.maxTime $\gets$ bmapElem.startTime + slaTable[bmapElem.vnoId].maxLatency
        \EndFor
    \EndFor
\EndProcedure

\Procedure{SortBmaps}{bmaps, nChannels, onuTunedChannelMap, slaBreachTable}
    \State Initialize BmapsChannel
    \For{$i \gets 0$ to $nChannels - 1$}
        \For{$\text{elem} \in \texttt{BmapsChannel}[i]$}
            \State $\text{onuId} \gets \text{elem.onuId}$
            \State $\text{channel} \gets \texttt{onuTunedChannelMap}[\text{onuId}]$
            \State Append $\text{elem}$ to $\texttt{BmapsChannel}[\text{channel}]$
        \EndFor
    \EndFor
    \Function{Compare}{elem1, elem2}
        \State slaBreach1 $\gets$ slaBreachTable[elem1.vnoId]
        \State slaBreach2 $\gets$ slaBreachTable[elem2.vnoId]
        \If{slaBreach1 $\neq$ slaBreach2}
            \State \Return slaBreach1 $<$ slaBreach2
        \ElsIf{elem1.maxTime $\neq$ elem2.maxTime}
            \State \Return elem1.maxTime $<$ elem2.maxTime
        \Else
            \State \Return elem1.size $<$ elem2.size
        \EndIf
    \EndFunction
    \For{$i \gets 0$ to $nChannels - 1$}
        \State $\texttt{BmapsChannel}[i] \gets \Call{MergeSort}{\texttt{BmapsChannel}[i], \texttt{Compare}}$
    \EndFor
    \State \Return BmapsChannel
\EndProcedure

\Procedure{UpdateSlaBreachTable}{BmapsChannel, slaBreachTable, numAllocsVno}
    \For{bmapChannel \textbf{in} BmapsChannel}
        \State Update schedTime and SLA breaches for each element
    \EndFor
\EndProcedure

\Procedure{StaticTWDM}{bmaps, slaTable, nChannels}
    \State Initialize state tables
    \While{\textbf{true}}
        \State \Call{CalculateMaxTime}{bmaps}
        \State BmapsChannel $\gets$ \Call{SortBmaps}{bmaps, nChannels, onuTunedChannelMap, slaBreachTable}
        \State \Call{UpdateSlaBreachTable}{BmapsChannel, slaBreachTable, numAllocsVno}
    \EndWhile
\EndProcedure
\end{algorithmic}
\end{algorithm}

\begin{figure}[htbp]
\centering
\includegraphics[width=.8\linewidth]{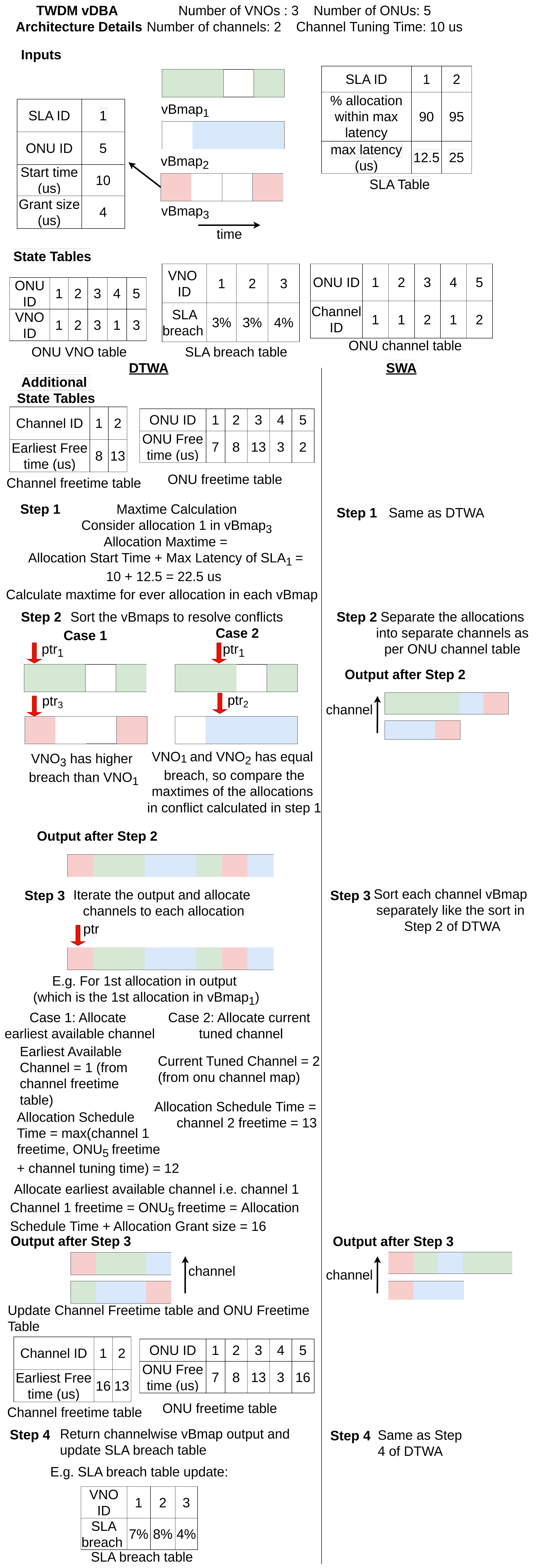}
\caption{DTWA Vs SWA Merging Algorithms}
\label{fig:dtwa_vs_swa}
\end{figure}

We illustrate the difference between the DTWA and SWA algorithms in terms of how we allocate wavelength to the allocations in figure \ref{fig:dtwa_vs_swa}. Here, we consider 3 VNOs and a total of 5 ONUs (we select a small number to simplify the explanation, but this can be extended to larger numbers of ONUs). Each ONU is associated to one of the VNOs as specified by the ONU VNO table. In this example, the number of transmission channels is 2 and the channel tuning time considered for all the ONU transceivers is 10 $\mu s$. The inputs to the algorithm are 3 Bmaps with the allocations in each Bmap represented by the same colour. For simplicity, here we consider a total of 4 allocations but it can be any number of allocations that can fit within a specific time window depending on the time window size and the network line capacity. Each allocation contains information like the SLA Id, ONU Id, start time and the grant size of the allocation. We have considered both the start time and grant size to be in $\mu s$. In our example, we just show the allocation information for the first allocation in $Bmap_3$ as we will explain only 1 iteration of allocation conflict resolution and wavelength allocation. Both algorithms maintains 2 state tables namely SLA breach table, which contains flow level SLA breach information so far for all the VNOs, and a ONU channel table to map the current tuned channel on the ONU transceivers. DTWA also maintains 2 additional tables namely a channel freetime table, which tracks the earliest time when the wavelength would be free again to transmit another allocation, and a ONU freetime table which tracks the earliest time when the ONU would be free again to transmit another allocation. In step 1 for both DTWA and SWA, as explained in code lines 4-7 in both DTWA\ref{alg:dtwa} and SWA\ref{alg:swa}, we calculate the maxtimes for all the allocations in each Bmap iteratively (in figure \ref{fig:dtwa_vs_swa} we just show the maxtime calculation for the first allocation of $Bmap_3$). Step 2 differs in DTWA and SWA. In case of DTWA, as explained in code lines 8-19 algorithm~\ref{alg:dtwa}, we append all the Bmaps and sort the appended Bmap which is explained as follows - for any 2 allocations in conflict, we first compare the SLA breach of the VNO flows to which the allocations belong respectively. There can be 2 possible cases as seen from the SLA breach table: case 1 where VNO $flow_3$ has higher SLA breach than VNO $flow_1$ in that case, the winning allocation will be that of the one belonging to $VNO_3$; case 2 where both VNO $flow_1$ and VNO $flow_2$ has the same breach, then we compare the maxtimes of both the allocations which we calculated in step 1, the winning allocation would be the one with earlier maxtime. The output of step 2 of DTWA is a sorted merged Bmap. For SWA, in step 2 we separate the allocations into separate channel Bmaps using the ONU channel table (code lines 9-14 algorithm~\ref{alg:swa}). All the allocations that will be transmitted in a particular channel are appended together. Hence, the output in this case are a single Bmap for each channel (in our case it is 2). The step 3 in DTWA is wavelength allocation to each of the sorted Bmap allocations (code lines 25 to 34). There can be 2 cases to consider - transmit the next allocation in the same wavelength as the ONU is already tuned to, or transmit at the earliest available wavelength. In figure \ref{fig:dtwa_vs_swa}, we have explained the cases illustratively for the first allocation of $Bmap_3$. The allocation belongs to $ONU_5$ which is already tuned in channel 2. In case 2, the allocation will be scheduled when channel 2 is free for transmission again after 13 $\mu s$. But in case 1, if we choose the earliest available wavelength (channel 1 as seen from the Channel Freetime table), the allocation scheduled time will be the maximum of channel 1 freetime and the sum of $ONU_5$ freetime and channel tuning time which is 12 $\mu s$. Hence, it is always optimal to switch to wavelength 1 for $ONU_5$ to transmit this allocation. Then we update the ONU channel table with the new channel (channel 1) just allocated to this ONU, the ONU freetime table and the channel freetime table which is the sum of the allocation schedule time and allocation grant size (16 $\mu s$). The output after this step in DTWA is the channelwise Bmap. In step 3 in SWA, we sort each channe-lwise Bmap using the sort function that we described in step 2 of DTWA (code lines 15-26 algorithm~\ref{alg:swa}). In step 4 in both DTWA and SWA, we calculate the new SLA breach for each of the VNO flows and update the SLA breach table (code lines 35-37 in algorithm~\ref{alg:dtwa} and code lines 27-29 in algorithm~\ref{alg:swa}).
 
\section{Simulation Setup for our Proposed Algorithm}
\label{experiment}
\begin{figure}[H]
\centering
\includegraphics[width=1\linewidth]{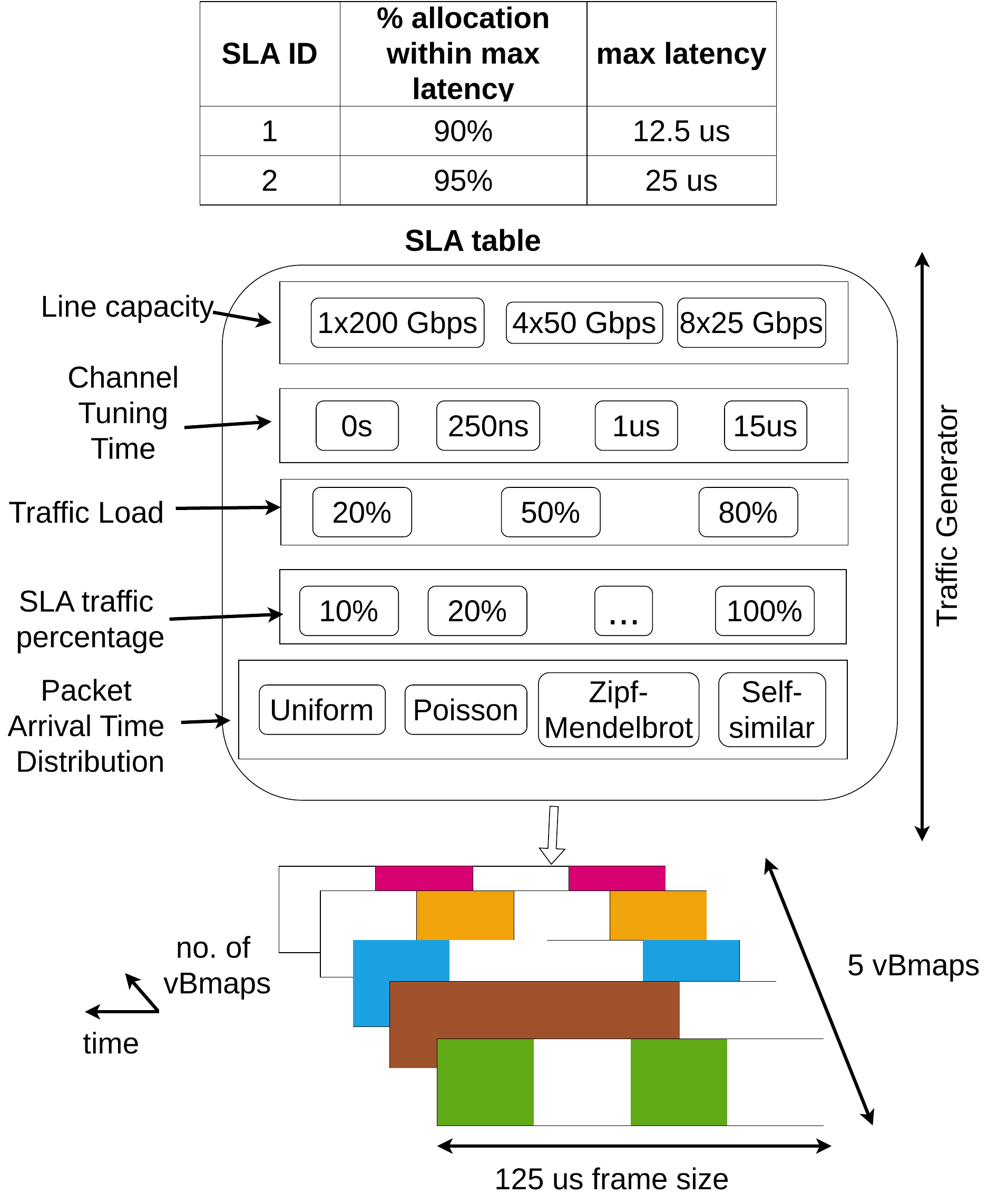}
\caption{Simulation Experiment}
\label{fig:simulation_exp}
\end{figure}

\begin{figure}[H]
\centering
\includegraphics[width=1\linewidth]{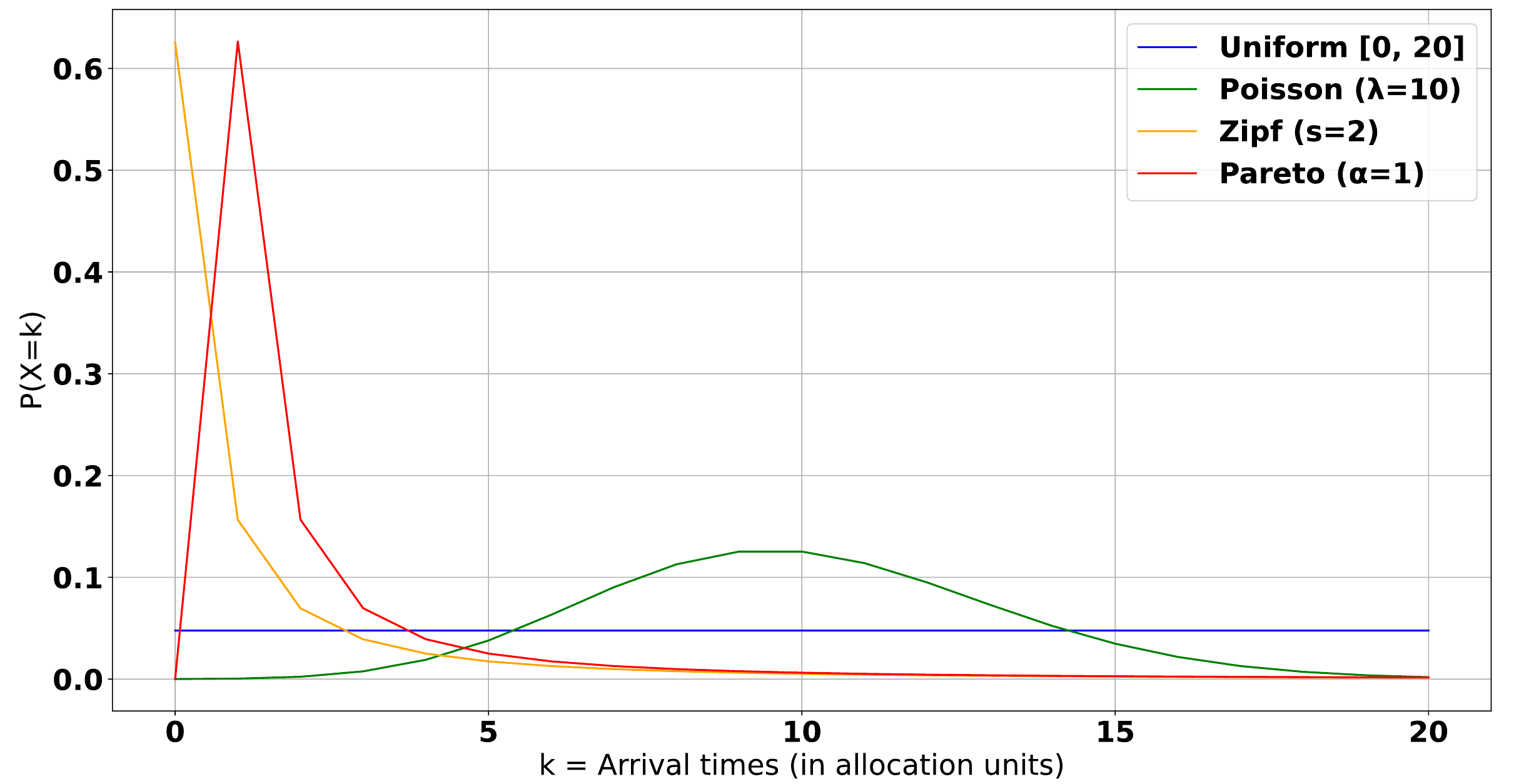}
\caption{Normalized Traffic Arrival Rate Distributions}
\label{fig:norm_distributions}
\end{figure}

We carried out a network traffic simulation to evaluate the performance of our merging algorithms (~\ref{alg:dtwa}~\nameref{alg:dtwa} and ~\ref{alg:swa}~\nameref{alg:dtwa}). The network simulation experiment is shown in figure \ref{fig:simulation_exp}. In order to set up the simulation environment, we generate input BMaps from different VNOs. The allocation load on the shared PON was considered for 20\%, 50\% and 80\% of the total upstream capacity (i.e., 200Gb/s across all channels considered). For each of these allocation loads, we then varied the percentage allocated to SLA-driven flows from 10\% to 100\% of the total load (the remaining part is allocated to best effort flows). We consider 5 VNOs (each one generating a virtual BMap every frame (125 $\mu s$)).
We consider a total of 64 ONUs, reflecting the typical split ratio used in both current and emerging PON deployments. These ONUs are evenly distributed among the 5 VNOs, such that four VNOs are assigned 13 ONUs each, and one VNO is assigned 12 ONUs. To avoid any bias in the VNO-to-ONU assignment, the mapping is randomized across simulation runs.
For our experiment, 2 types of SLAs have been chosen: one requiring 90\% compliance with a latency target of 12.5 $\mu s$ and one requiring 95\% compliance with a latency target of 25 $\mu s$. This type of QoS cannot be accomplished with classical priority, because the stricter requirement for compliance and latency are alternated between the two SLAs. Each bandwidth map has a set of allocations, the burst sizes were generated from a uniform distribution with average burst size set at 4\% of the total frame size (the same ONU is also allowed to provide multiple burst per frame).An empty time slot of 0.21 $\mu s$, is introduced between allocations to account for guard time between upstream transmissions. The minimum and maximum burst sizes were fixed such that the guard times account for 25\% and 3\% of the burst size respectively.
The traffic arrival times were generated using 4 different probability distributions - uniform, Poisson, Zipf-Mendelbrot and self-similar distribution. Note that we have chosen different distributions for generating the allocation sizes and the allocation arrival times. This is because the allocation sizes depend on the type of user applications but the arrival times of the allocations depend on the network load, user behavior, and the nature of data transmission. By evenly distributing the allocation sizes within a defined range, we can evaluate how well our merging algorithm performs when handling packets of various sizes. We used Pareto distribution which characterizes the self-similar distribution. The various distribution parameters are shown in figure \ref{fig:norm_distributions}. Note that here we normalize the distributions for the chosen range of arrival times so as to make sure the arrival times outside that range is unlikely. The arrival times of the BMap allocations are randomly chosen from a range of 0 to 20 allocation units for each of the distributions. One allocation unit is the smallest upstream grant that can be allocated to one ONU and for our system, we consider a size of 160 bytes or approximate duration of 0.05 $\mu$s. The choice of the range is chosen such that the average arrival time (for uniform and Poisson distribution) is 10\% of the average allocation size. This is to make sure that the generated bandwidth maps are not sparse in nature and there is a high probability of allocation conflicts among them. For the uniform distribution, the arrival times are randomly but uniformly chosen from the 0-20 allocation units range with an expected arrival rate of 10 allocation units. The expected arrival rate ($\lambda$) for the Poisson distribution is chosen as 10 allocation units, same as the case of uniform distribution. For the Zipf-Mendelbrot distribution, the exponent characterizing the distribution's tail is $s = 2$. For the Pareto distribution, we have chosen the shape parameter $\alpha = 1$. This ensures that both the Zipf-Mendelbrot and Pareto distributions are highly skewed to the right, meaning shorter arrival times are more likely with rare long delays in between. Each ONU has one transceiver, which is tunable on any of the available channels, and constrained by a tuning time (for the case of the DTWA algorithm (\ref{alg:dtwa})), which is a simulation parameter. As mentioned above, the system under investigations, are 8 x 25Gb/s, 4 x 50Gb/s and 1 x 200Gb/s channels. The tuning times considered span from negligible to 250 ns, 1 $\mu s$ and 15 $\mu s$. The experiment was run for 1000 time frames and the average number of SLA breaches is recorded. The experiments were conducted on a Dell Precision 3660 workstation equipped with a 13th Gen Intel\textsuperscript{\textregistered} Core\textsuperscript{TM} i9-13900K CPU. The processor has 24 cores (32 threads) with a base clock speed of 3.00\,GHz and a maximum turbo frequency of up to 5.80\,GHz. The system includes 64\,GB of DDR5 RAM (2\,×\,32\,GB modules) configured at 4400\,MT/s. All experiments were executed on Ubuntu 22.04.5 LTS with Linux kernel 6.8.0-49-generic.

\section{Results and Discussion}
\subsection{SLA performance for variable channel systems and tuning times}
\begin{figure}[H]
    \centering
    \includegraphics[width=0.5\textwidth]{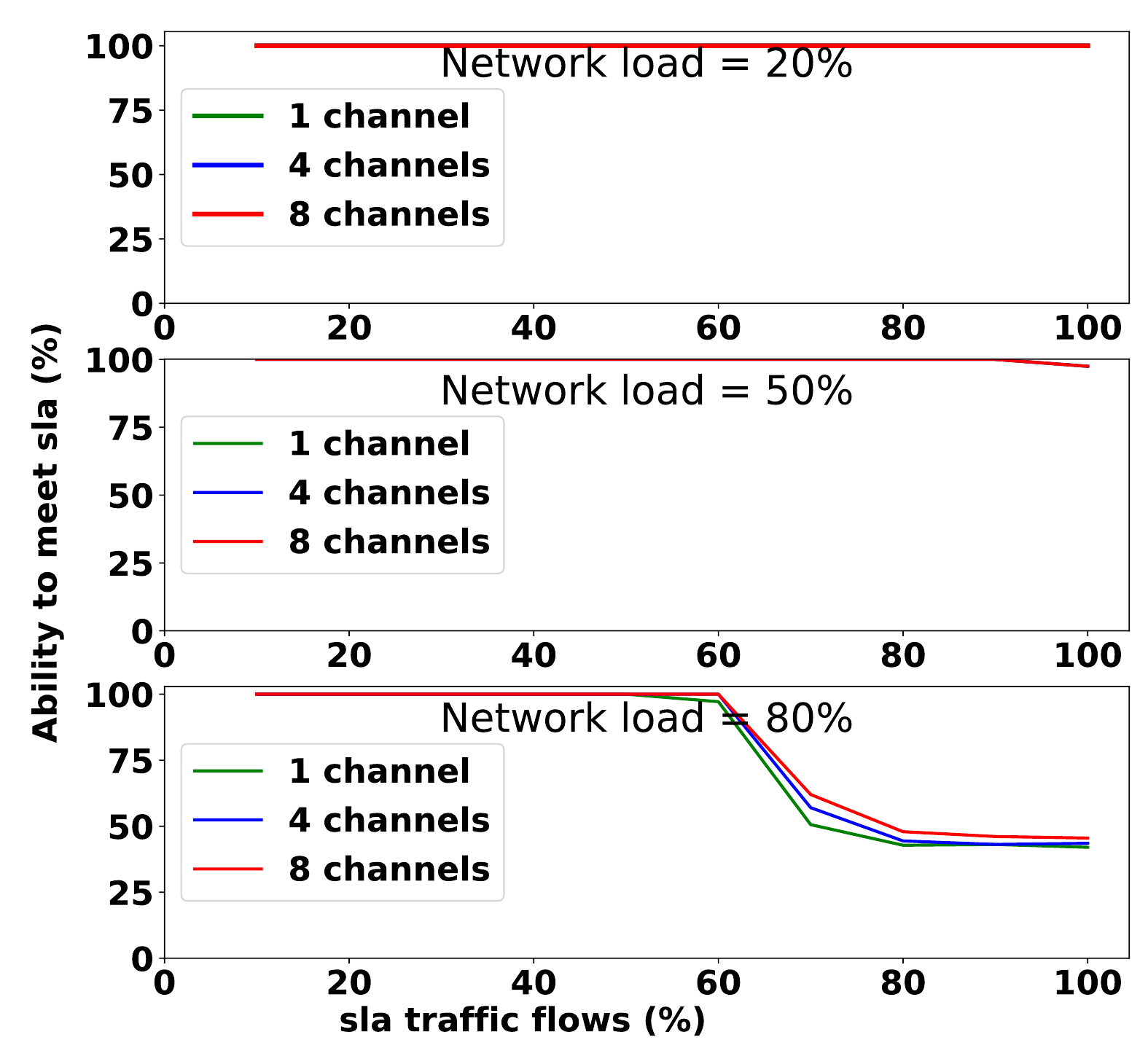}
    \caption{DTWA loading with Tuning time 0s}
    \label{fig:0s}
\end{figure}

\begin{figure}[H]
    \centering
    \includegraphics[width=0.5\textwidth]{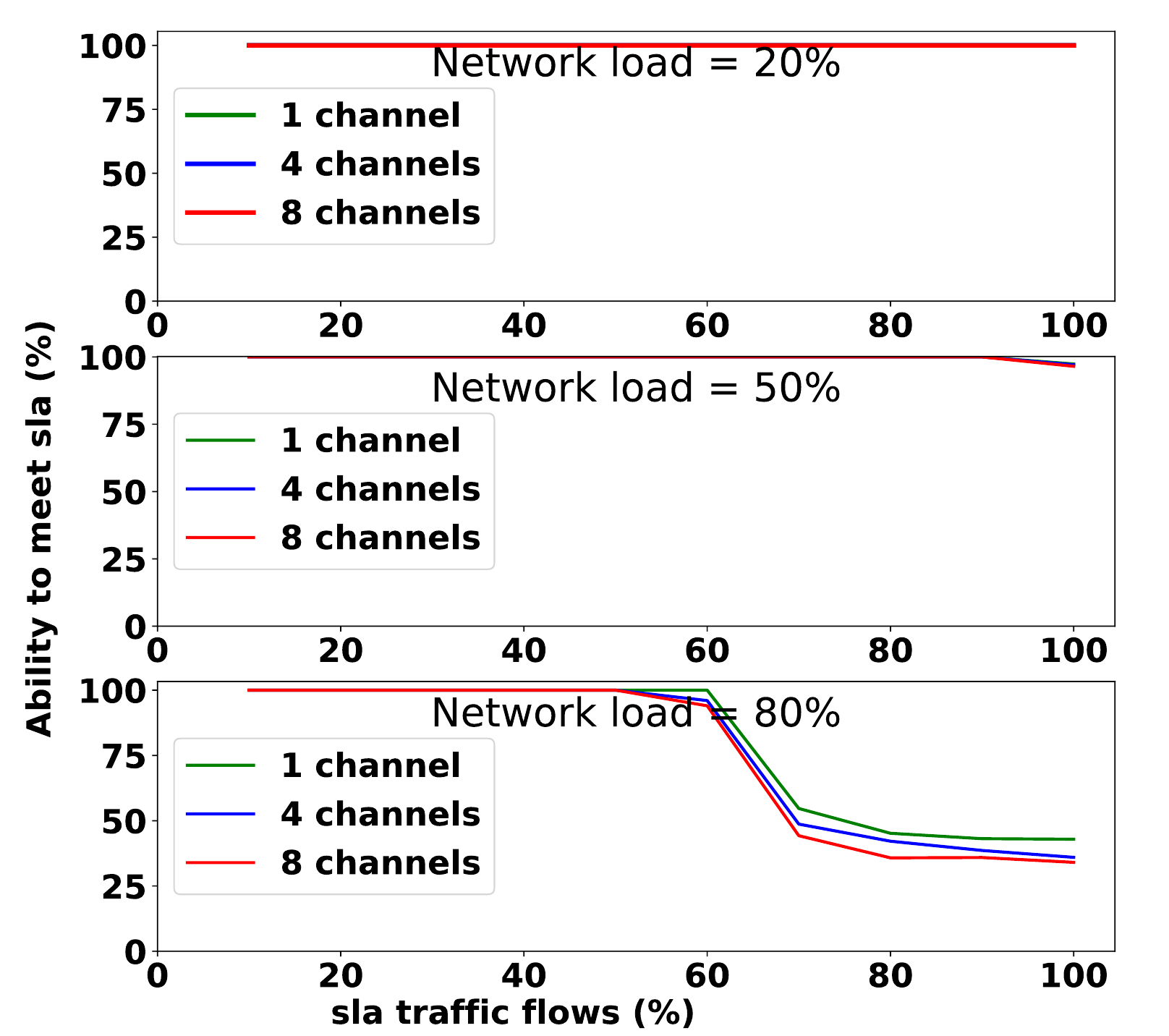}
    \caption{DTWA loading with Tuning time 250 ns}
    \label{fig:250ns}
\end{figure}

\begin{figure}[H]
    \centering
    \includegraphics[width=0.5\textwidth]{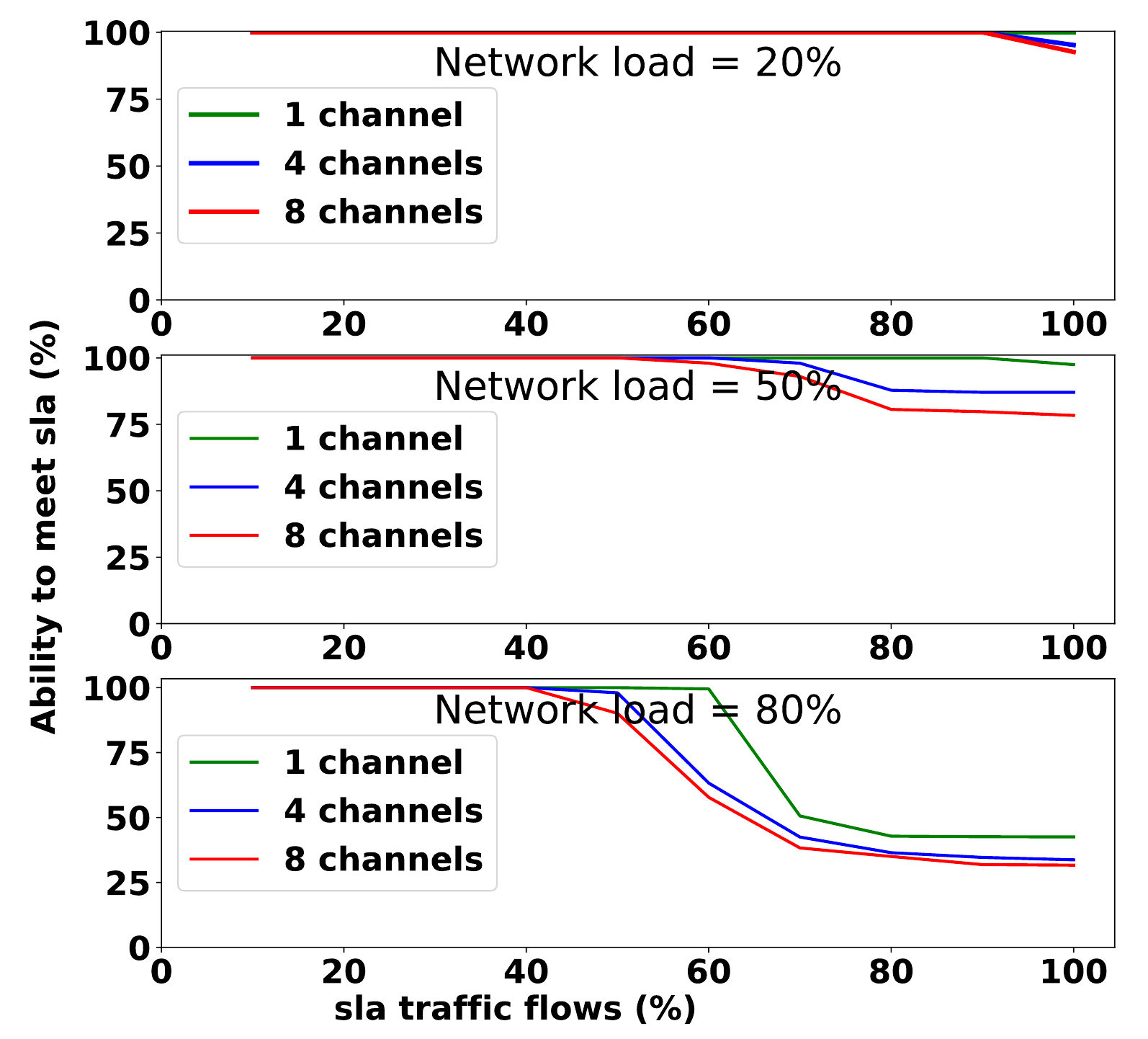}
    \caption{DTWA loading with Tuning time 1$\mu$s}
    \label{fig:1us}
\end{figure}

\begin{figure}[H]
    \centering
    \includegraphics[width=0.5\textwidth]{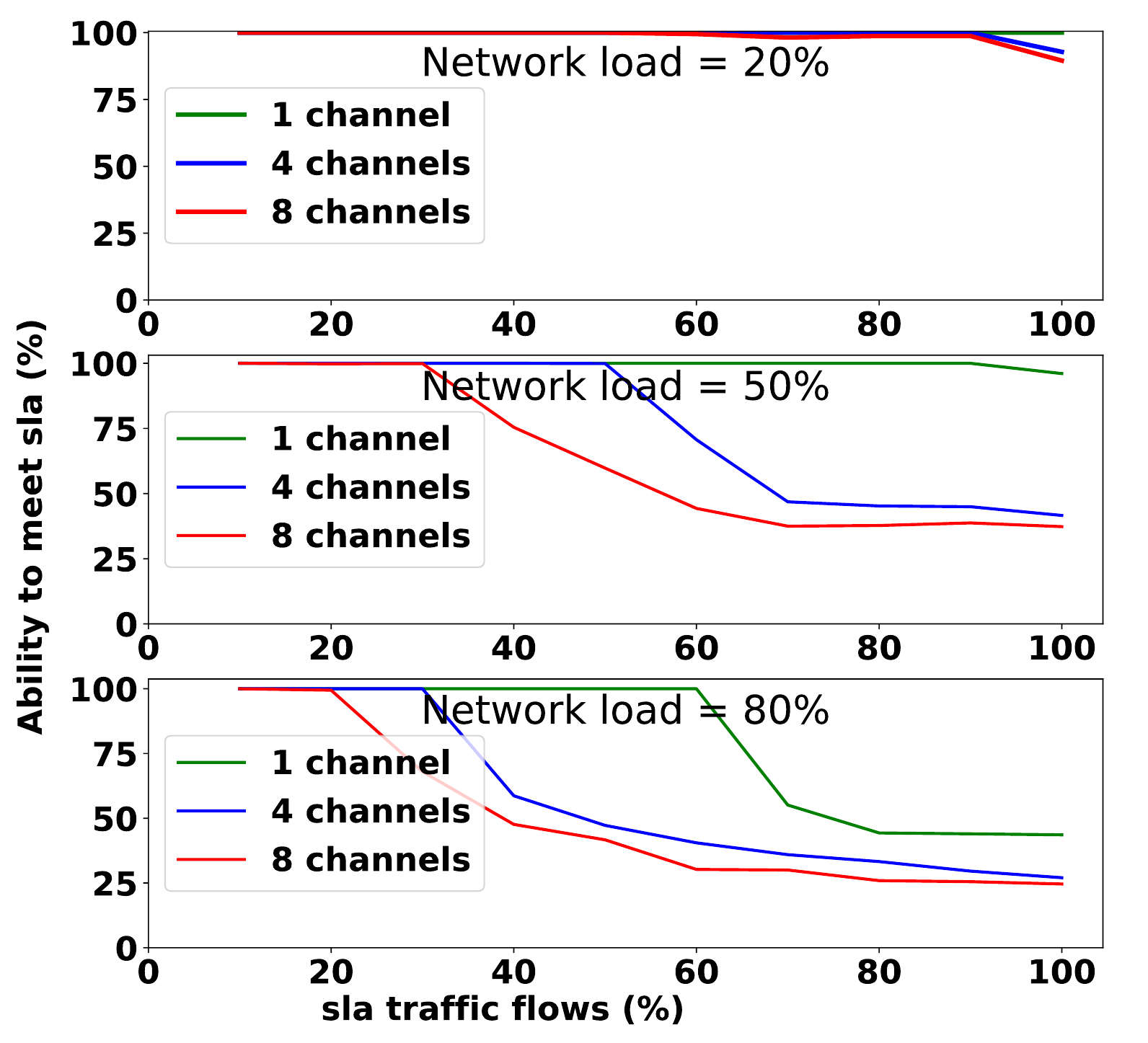}
    \caption{DTWA loading with Tuning time 15$\mu$s}
    \label{fig:15us}
\end{figure}

The results we report in this section demonstrate the system's capability to comply with SLAs as a function of the proportion of flows requiring SLA guarantees. Each figure includes three curves corresponding to PON loads of 20\%, 50\%, and 80\% of total system capacity. Different plot colors represent different channel configurations, specifically 8x25G, 4x50G, and 1x200G transceivers. The performance variation due to different transceiver tuning times is illustrated in \cref{fig:0s,fig:250ns,fig:1us,fig:15us}, and the scheduling behavior is evaluated using the algorithm described in \cref{alg:dtwa}.
The plot in Fig~\ref{fig:0s} reports the ability to meet SLA, when the tuning time is negligible, providing an upper bound on performance as we compare the effect of different tuning times. Under this setting, we observe near-perfect SLA satisfaction for all PON loads up to 50\%, with only a minor degradation for SLA traffic fractions exceeding 90\%. At 80\% PON load, the system begins to show performance degradation: the 8x25G and 4x50G configurations experience SLA violations starting around 60\% SLA traffic, while the 1x200G configuration sees a drop starting at 50\%. This confirms that higher aggregate loads increase contention of slots between different Bmap flows and reduce scheduling flexibility in terms of SLA compliance. Notably, the multi-channel configurations (8x25G and 4x50G) outperform the single-channel system (1x200G), even when tuning time is negligible. This occurs because the PON burst overhead is of the order of a few hundred ns, independent of the data rate, thus the drop in throughput is proportionally larger for higher rate channels as they contain higher number of allocations within the same Bmap time frame.

Figures~\ref{fig:250ns}, \ref{fig:1us}, and \ref{fig:15us} analyze the impact of increasing tuning time on SLA compliance. As expected, SLA satisfaction degrades with higher tuning delays. This is because our merging algorithm progressively loses the ability to avoid scheduling delays by tuning to different channels. At a tuning time of 250~ns, the system still performs resonably well in terms of SLA compliance, but noticeable drops emerge for higher loads and higher SLA traffic ratios. At 1~$\mu$s and especially at 15~$\mu$s, the performance of 8x25G and 4x50G systems substantially declines, confirming the strong sensitivity of channel switching schemes to tuning delays.

Importantly, the 1x200G configuration remains unaffected across all tuning time scenarios. This is because it does not perform any channel tuning—the system operates a single fixed channel, thus incurring no switching overhead. While this setup avoids tuning penalties, it is still limited by higher burst overhead.

The results clearly highlight a fundamental trade-off: while multi-channel configurations provide better statistical multiplexing and finer granularity under ideal tuning conditions, their advantage diminishes or reverses under practical tuning constraints. From a system design perspective, these results suggest that for architectures with non-negligible tuning times (e.g., $\geq$250~ns), optimizing channel configuration and tuning policies is critical. Furthermore, these observations emphasize that SLA-aware algorithms must account not only for traffic demand but also for underlying hardware limitations, such as tuning delay, to ensure SLA compliance in high-load scenarios.

Overall, the analysis underscores the necessity of considering physical-layer constraints—especially tuning time—in the design and evaluation of multi-channel PON scheduling algorithms.

\subsection{SLA performance for different traffic distributions}
\label{SLA_performance_variable_traffic}
\begin{figure}[H]
    \centering
    \includegraphics[width=.8\linewidth]{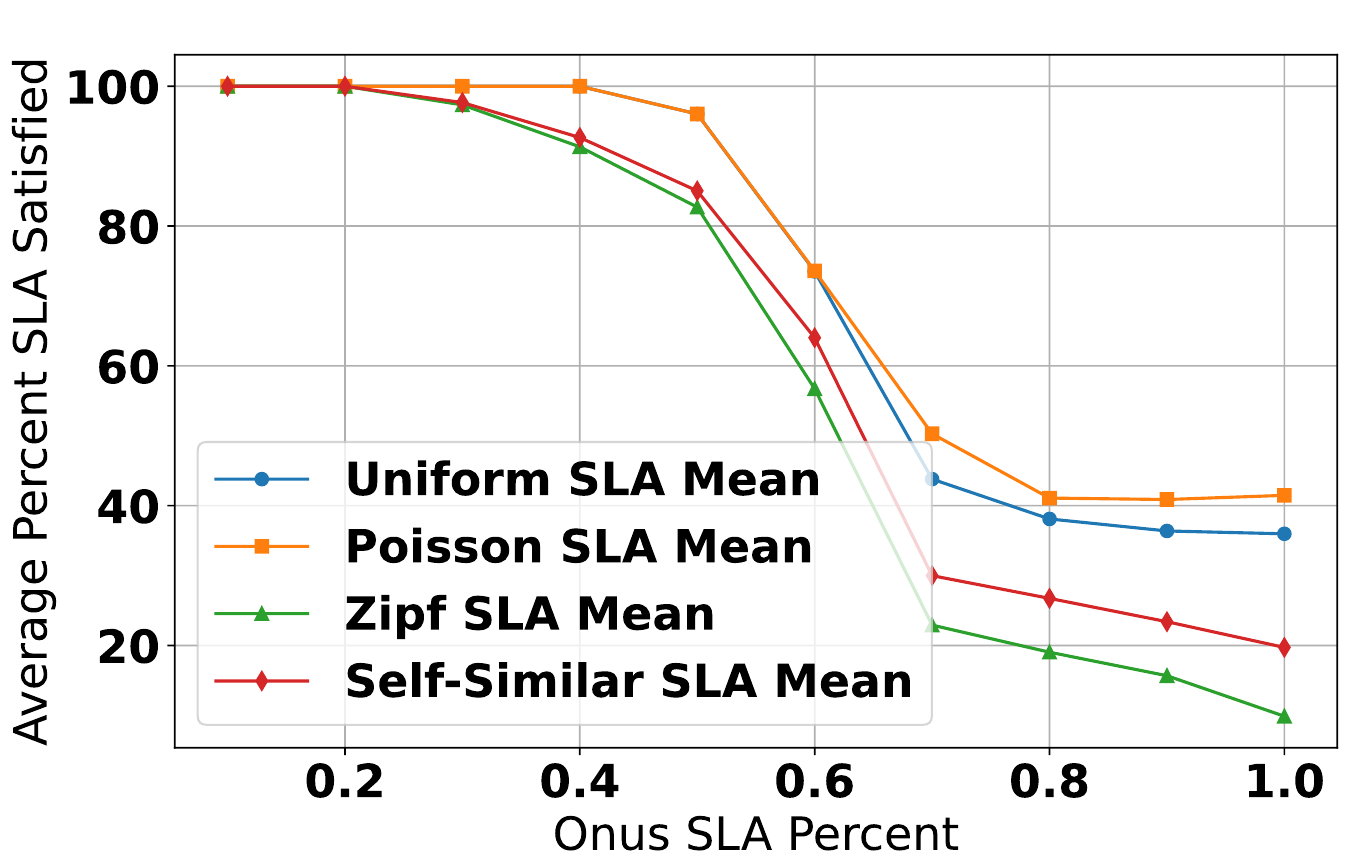}
    \caption{Average SLA performance of the DTWA algorithm for different traffic distributions}
    \label{fig:distribution_cmps}
\end{figure}

The impact of different traffic arrival distributions on the average ability to meet SLAs is shown in Fig.~\ref{fig:distribution_cmps}. To ensure that the results are not biased by specific tuning delays, channel capacities, or network loads, we average the SLA compliance values across all combinations of these parameters considered in the previous experiments. This gives a generalized view of how the traffic distribution alone affects the scheduling algorithm’s performance.

The results reveal that the system performs significantly better under uniform and Poisson traffic arrival distributions compared to Zipf-Mendelbrot and self-similar (Pareto) distributions. This can be attributed to the statistical properties of these distributions.

In both uniform and Poisson distributions, traffic arrivals are more evenly spaced across the allocation window. For the uniform case, each arrival time within the considered range is equally probable, resulting in a low probability of Bmap allocation conflict. In the Poisson distribution, the arrivals are probabilistically centered around a mean (here, 10 allocation units or 0.5~$\mu$s), which leads to even better temporal spreading due to lower variance and higher likelihood of arrivals near the mean. This means that the allocations within each BMap are temporally dispersed, reducing the chance of allocation collisions and thereby improving SLA compliance. 

Moreover, the Poisson distribution slightly outperforms the uniform case. This is because, although both provide fairly symmetrical and moderate-variance arrival profiles, Poisson arrivals tend to cluster mildly around the expected value. This mild clustering minimizes high-variance "gaps" and "bursts" that are more typical in the uniform case, which can lead to variable congestion periods and transient SLA violations.

Conversely, for Zipf-Mendelbrot and self-similar distributions, the probability mass is skewed toward shorter inter-arrival times. These right-skewed distributions lead to temporal clustering of allocation events, which increases the likelihood of contention for transmission windows. In particular, the Zipf-Mendelbrot distribution exhibits a heavier tail than the self-similar distribution, resulting in more extreme short-term bursts. As a result, the performance of DTWA under Zipf-Mendelbrot traffic is significantly degraded due to a high incidence of overlapping requests within a BMap, which limits the scheduler’s flexibility to resolve conflicts within strict SLA deadlines.

These findings emphasize that the effectiveness of the DTWA algorithm is highly dependent on the temporal characteristics of traffic. Distributions that offer more evenly spaced arrivals (with moderate or low variance) enable better scheduling opportunities and lower contention, whereas bursty or highly skewed traffic patterns result in greater conflicts and higher SLA violations.

\subsection{SLA performance comparison between DTWA and SWA}
\begin{figure}[H]
    \centering
    \includegraphics[width=0.5\textwidth]{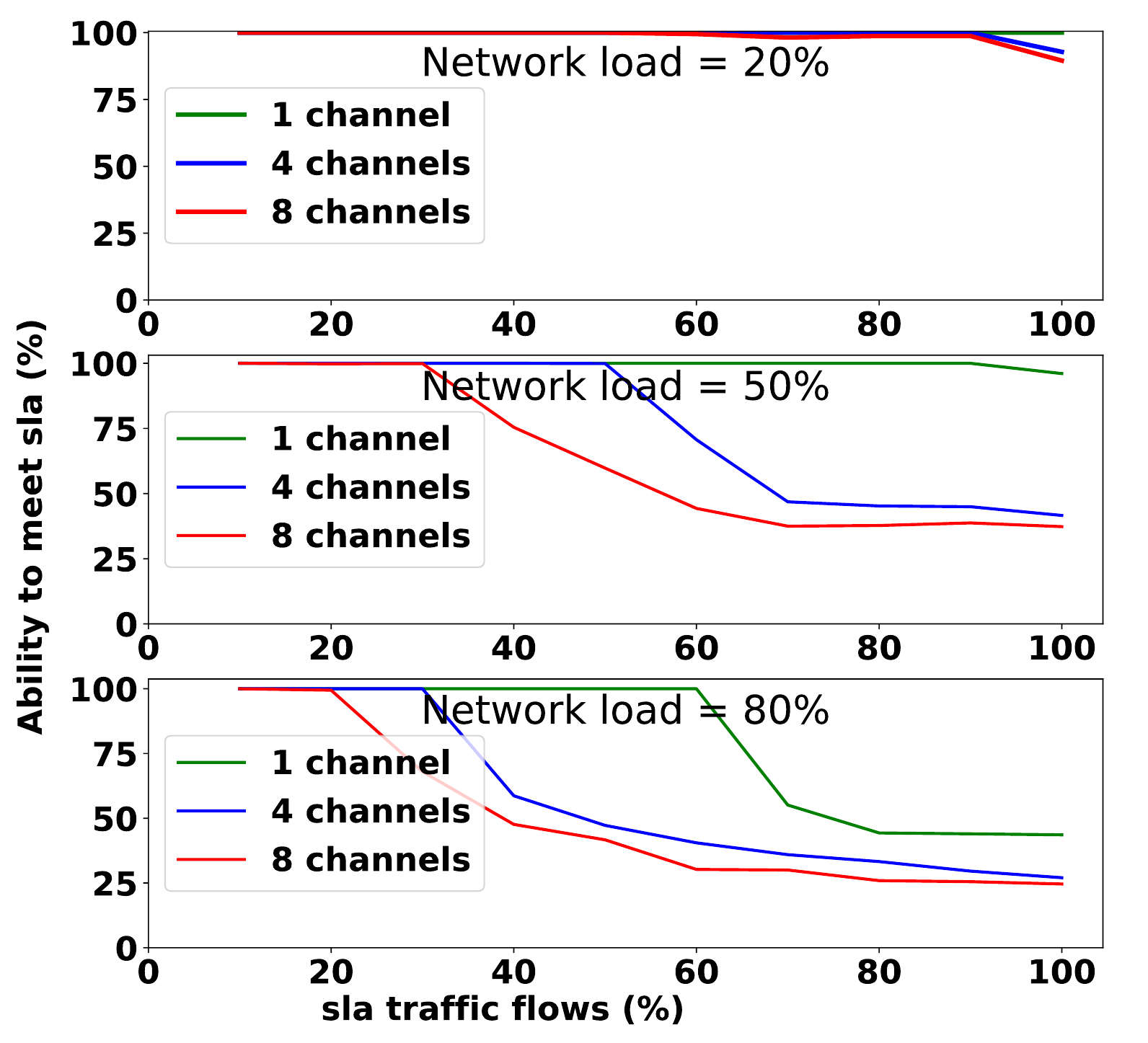}
    \caption{SLA Performance of the SWA algorithm}
    \label{fig:sla_static_twdm}
\end{figure}

We also examined the SLA performance of the static wavelength assignment (SWA) algorithm (\cref{alg:swa}) under a uniform traffic distribution, as reported in Fig.~\ref{fig:sla_static_twdm}. We compare these results with those of DTWA in the case of a 15~$\mu$s tuning delay (Fig.~\ref{fig:15us}). We observe that the performance of SWA is comparable to DTWA when tuning time is high. This is expected, as the advantage of dynamic tuning diminishes with increased tuning delays. At 15~$\mu$s, the cost (in terms of idle time and scheduling delay) of switching wavelengths outweighs the potential gain in flexibility. Therefore, assigning a fixed channel per ONU, as in SWA, becomes more effective—or at least equally viable—when tuning incurs high latency penalties.

It is important to note that our experiments assume equal average load across all ONUs, which helps isolate the impact of short-term allocation strategies. By focusing on short timescale capacity scheduling, we capture the responsiveness of the algorithms to transient demand variations and contention patterns driven by the arrival distribution, rather than long-term load imbalance.

These results provide valuable insights for the design of bandwidth allocation algorithms in dynamic TWDM-PONs. They suggest that traffic-aware scheduling strategies should account not only for load but also for traffic burstiness, and that static channel allocation may be a better alternative in scenarios with hardware-imposed tuning limitations.

\subsection{SLA performance comparison between DTWA and CPLEX}
\label{results_dtwa_cplex}
\begin{figure}[H]
    \centering
    \includegraphics[width=0.5\textwidth]{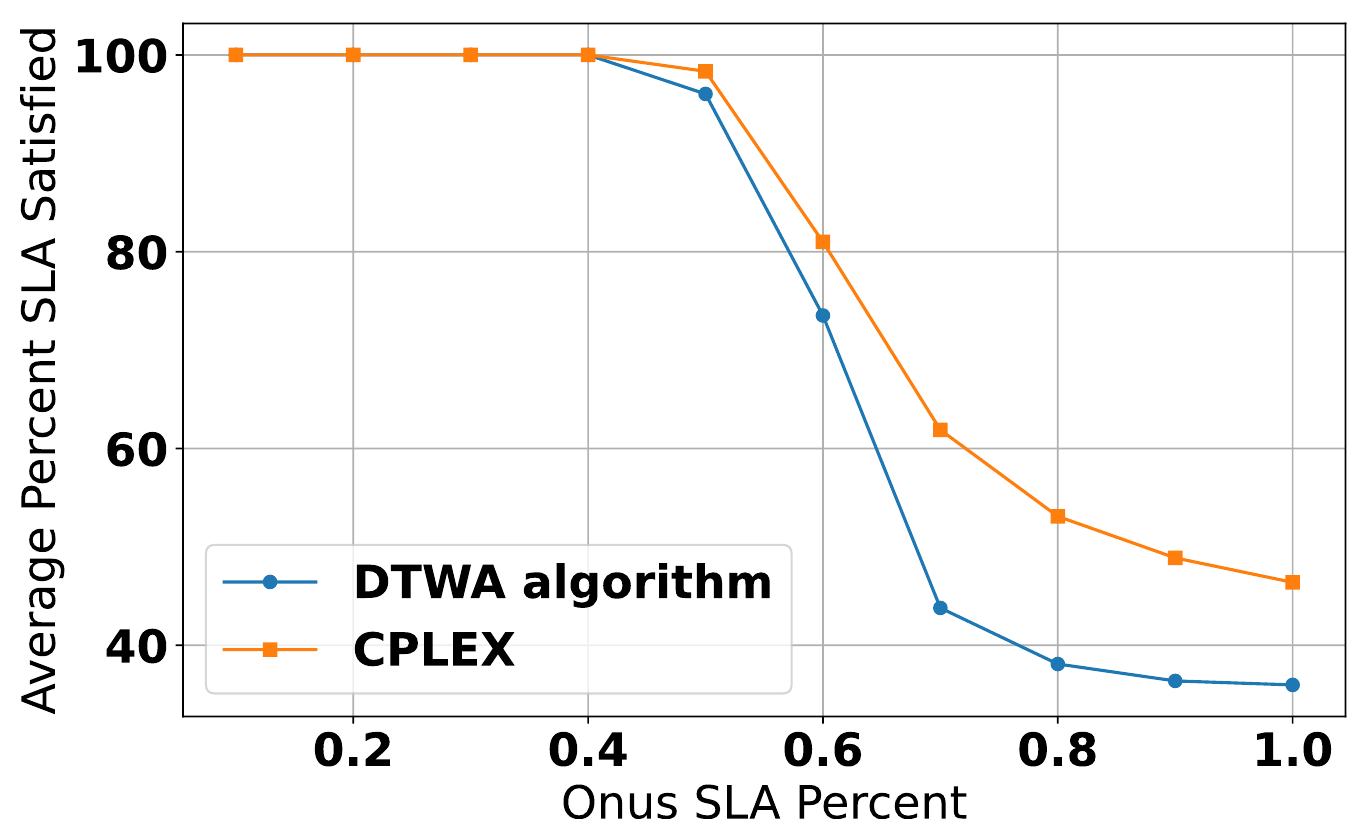}
    \caption{Average SLA Performance of DTWA and CPLEX}
    \label{fig:cplex_dtwa_cmp}
\end{figure}

Finally, we compare our DTWA algorithm against the optimal MIP solution provided in section \ref{section2} (through equations \eqref{eq:objective}, to \eqref{eq:constraint3}) as shown in figure \ref{fig:cplex_dtwa_cmp}. 
We fix the channel tuning time at 250 ns and the network load at 80\% of the maximum possible load for a total aggregated line capacity of 200 Gbps and solve the MIP problem and generate an optimal solution using the IBM ILOG CPLEX Optimizer module. Then we average the SLA performance of the results obtained using CPLEX and DTWA across the different channel capacities. The comparison plot indicates that the performance of our DTWA algorithm is close to optimal until the SLA traffic percentage does not exceed 60 \%, beyond that we can observe a significant drop in the performance of DTWA from the optimal CPLEX solution. 
This is because of fundamental differences in decision-making between DTWA and CPLEX. The search space of the Bmap merging problem is one of the many permutations of all the allocation slots from all the Bmaps. To prune the search space, CPLEX uses a combination of state-of-the-art mathematical optimization algorithms namely Branch and Cut \cite{cplex_branch_cut}, Presolve and Heuristics \cite{cplex_mip_heuristics}, Cutting Planes \cite{cplex_mip_cuts} and Dynamic Search Strategies \cite{cplex_dynamic_search}, whereas DTWA is a real-time greedy algorithm which prioritizes resolving contention based on immediate SLA breaches without performing a full search over potential future outcomes. As SLA traffic increases, contention for slots becomes more intense, and DTWA's heuristic becomes less effective at finding the globally optimal allocation.

\subsection{Algorithm execution time}
\begin{figure}[H]
    \centering
    \includegraphics[width=0.5\textwidth]{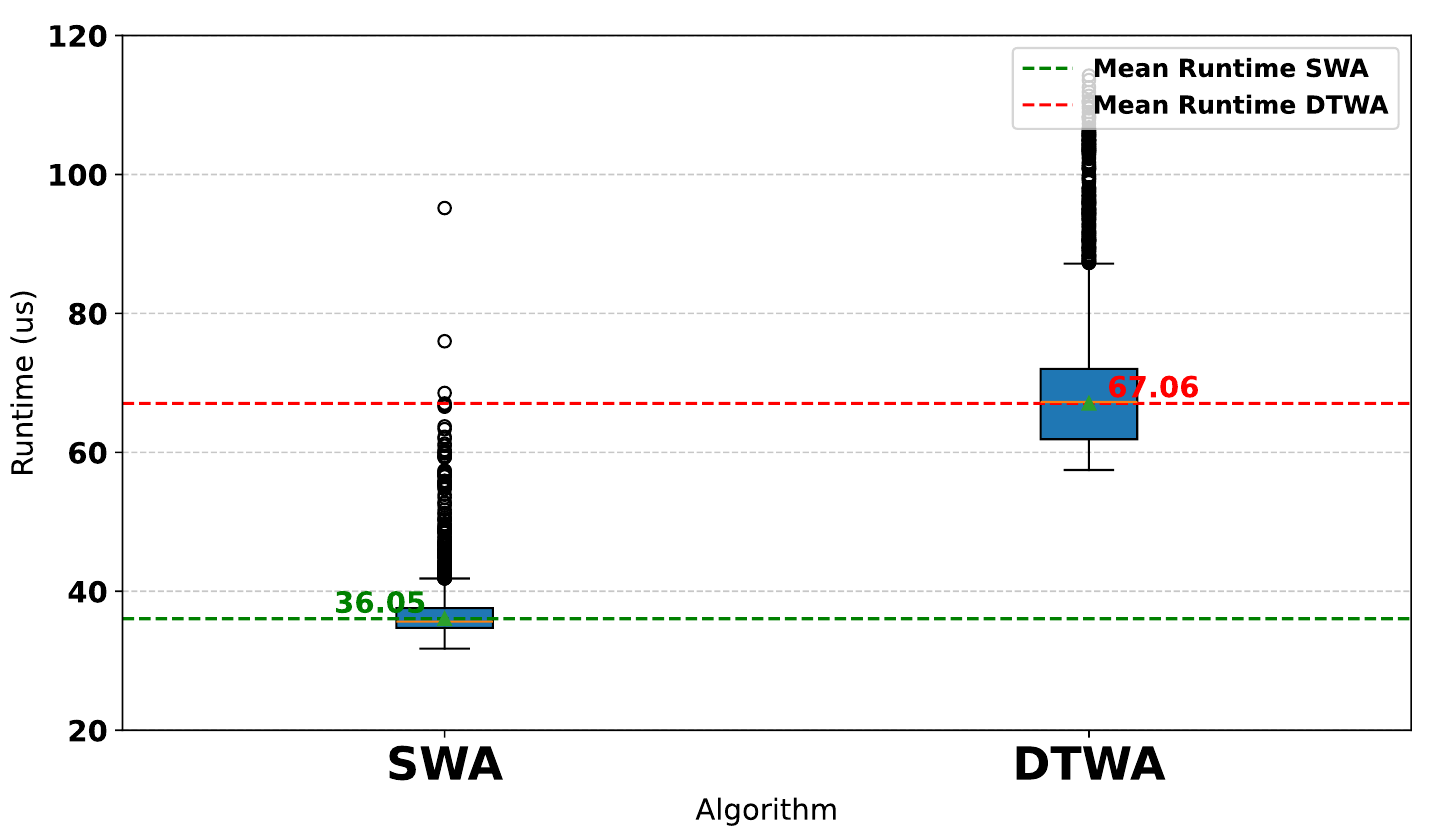}
    \caption{Boxplot for the runtimes of the algorithms}
    \label{fig:runtime_boxplots}
\end{figure}

\begin{figure}[H]
    \centering
    \includegraphics[width=0.5\textwidth]{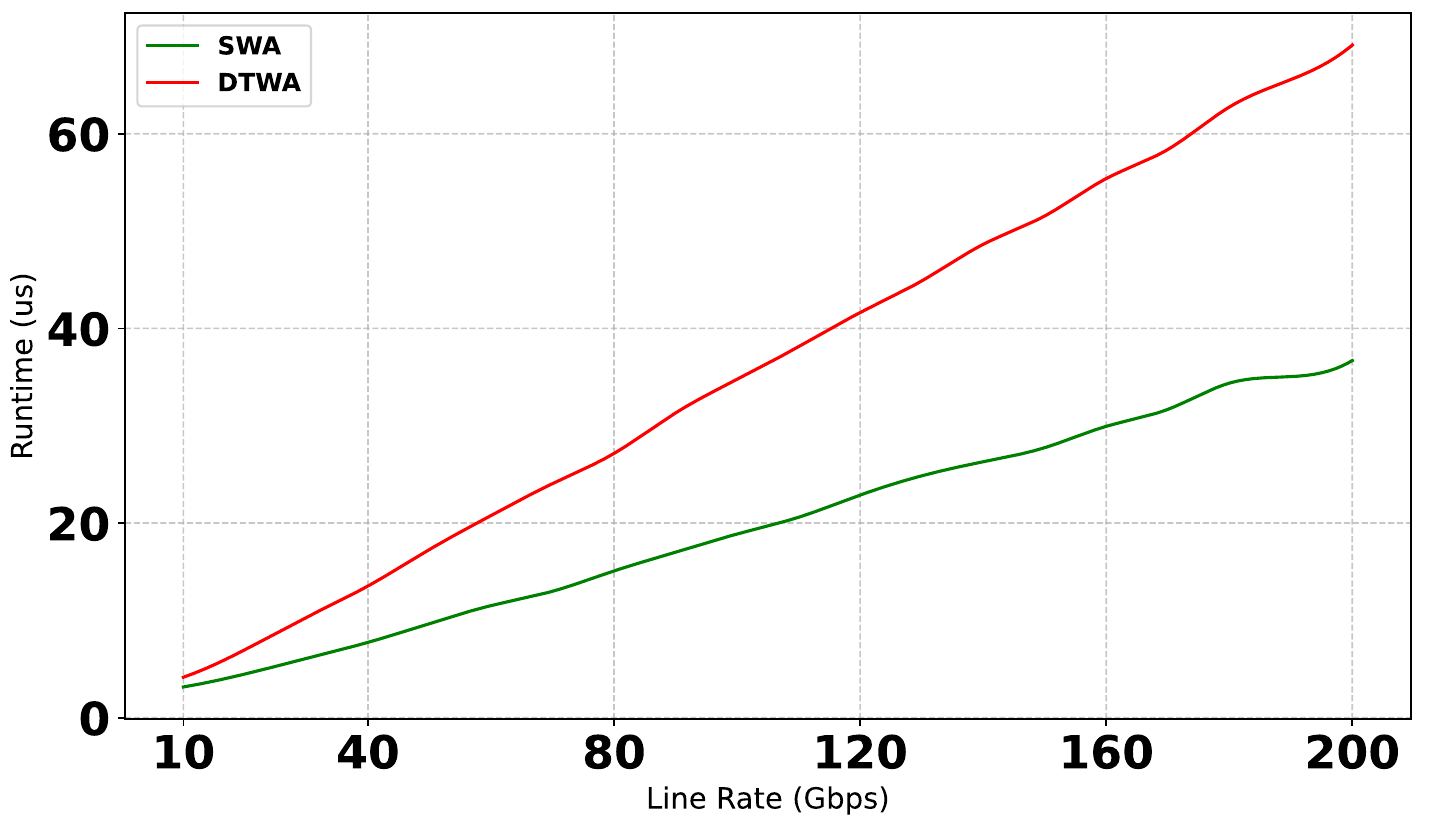}
    \caption{Runtimes for different line capacities}
    \label{fig:line_rate_runtimes}
\end{figure}

In this section we report the results of the run time analysis of the two proposed algorithms. Since the merging algorithm needs to operate in real time, it is important to assess their execution time. The plot in figure \ref{fig:runtime_boxplots} shows the profiled runtimes of the algorithms run for 1000 occurrences for a 200Gbps aggregated network capacity system. In average, SWA is $1.9$ times faster than DTWA. The interquartile range (IQR) of SWA is smaller than that of DTWA, which means SWA has a more predictable and consistent runtime. Each step of allocation conflict resolution is accompanied by selection of an ideal wavelength for transmission in DTWA whereas in SWA, the choice of wavelength is avoided as the already tuned wavelength is allocated again to the ONU for transmission (\cref{alg:dtwa,alg:swa}). The additional system instructions executed for wavelength selection in DTWA leads to higher variance in its runtimes. 
We also vary the total line capacity of the system and measure the runtime for both algorithms, as shown in Fig.~\ref{fig:line_rate_runtimes}. The runtimes are directly proportional to the line capacity, because the total number of allocations ($n$), combined across all BMaps, is directly proportional to the amount of data that needs to be scheduled. Both algorithms depend on a resource-intensive merge sort operation, whose time complexity is $\mathcal{O}(n \log n)$, where $n$ represents the number of elements to be sorted. For the more complex DTWA algorithm, the slope is larger than that of SWA. This means that the additional operations introduced while sorting the allocations in DTWA add an extra latency component during the execution of the algorithm.

In addition to the line rate, the value of $n$ also scales with the number of Virtual Network Operators (VNOs), since each VNO contributes one bandwidth map per frame. A larger number of VNOs increases the total number of allocations to be processed. Furthermore, the number of allocations within each bandwidth map generally increases with the number of Optical Network Units (ONUs) per VNO. However, this increase is not strictly linear, as individual ONUs can issue multiple allocation requests within the same 125~$\mu$s frame under high traffic load. This introduces additional variability and burstiness in the traffic, making the merging process more computationally demanding. As a result, increases in the number of VNOs and ONUs also contribute to the growth of $n$, and the $\mathcal{O}(n \log n)$ complexity can become a performance bottleneck under such high-load or bursty conditions.

To further support our algorithm's efficiency, we compared DTWA against an optimal solution obtained using the IBM ILOG CPLEX Optimizer. For this comparison, we fixed the tuning time at 250\,ns and the network load at 80\%, evaluated over all three channel capacities, 10 different SLA traffic percentages (from 10\% to 100\%), and 1000 iterations per configuration, as described in Section~\ref{results_dtwa_cplex}~\nameref{results_dtwa_cplex}. The total execution time for all CPLEX simulations was 55,520 seconds. On average, computing a single merged bandwidth map using CPLEX takes approximately 1.85 seconds, which is several orders of magnitude slower than the DTWA algorithm that executes within a few microseconds.

Overall, we can see that a software-based implementation might be feasible (i.e., an additional latency of less than 10~$\mu$s) for rates around 50\,Gb/s in a single-wavelength static system, while the dynamic TWDM case would likely require hardware acceleration—especially considering that practical systems may operate with multi-channel configurations at 100\,Gb/s and beyond. 

\section{Challenges and Future Work}
\subsection{Mitigating SLA Variance Across Traffic Patterns in PON}
As discussed in Section~\ref{SLA_performance_variable_traffic}~\nameref{SLA_performance_variable_traffic}, we observed that SLA compliance was influenced by the nature of traffic distributions in the network. While our proposed merging algorithm achieved high SLA compliance under uniform traffic like the Uniform and Poisson distributions, its performance reduced under bursty or self-similar patterns, such as those modeled by Pareto or Zipf-Mendelbrot distributions.

Recent studies have underscored the inherent challenges in accurately predicting network traffic at fine time scales. These challenges arise due to the non-stationarity, high variability, and burstiness of traffic in modern broadband and access networks. For instance, bursty traffic often exhibits long-range dependence and heavy-tailed behavior, making fine-grained forecasts of packet arrival rates practically infeasible~\cite{jahnke2018finegrainednetworkflow}. Even deep learning models like LSTM and GRU fail to generalize under non-repeating traffic patterns and tend to regress to mean predictions~\cite{traffic_prediction_jahnke}.

While precise packet-level prediction remains an open challenge, it is considerably more feasible to perform high-level classification of traffic patterns. Classification of ONUs into categories such as \textit{uniform} and \textit{bursty} has been shown to be relatively stable over time and more robust to noise~\cite{upstream_bandwidth_allocation_frigui, burstiness_ouah}. Such classifications can be derived from historical traffic statistics, inter-arrival time distributions, or bandwidth request variability.

To mitigate the SLA performance variance, we propose two potential solutions based on traffic pattern classification:
\begin{enumerate}
\item \textbf{Distribution-Aware DTWA:} By classifying ONUs as \textit{bursty} or \textit{uniform}, we can allocate dedicated transmission windows for bursty ONUs using our DTWA algorithm. This could reduce contention during bandwidth allocation, particularly under high-load or bursty scenarios, and lead to improved SLA compliance.
\item \textbf{Map-Level Bandwidth Profile Classification:} Alternatively, we can perform online classification of ONUs requiring high bandwidth based directly on bandwidth map inputs. This bypasses traffic distribution inference errors but requires additional bandwidth map traversal and processing time. Despite the higher runtime, it offers a more accurate and adaptive bandwidth map merging strategy for real-time deployments.
\end{enumerate}
These classification-based adjustments serve as practical enhancements to our merging algorithm and offer a promising direction for future work. They strike a balance between runtime feasibility and improved performance robustness under diverse traffic conditions.

\subsection{O-RAN Fronthaul Support with PON Integration}
We mentioned in Section~\ref{introduction}~\nameref{introduction} that virtualized PONs (vPONs), due to their cost-effectiveness and high scalability, have been considered to support O-RAN fronthaul deployments. We aim to integrate our bandwidth map merging algorithm within such a vPON architecture supporting O-RAN fronthaul. Specifically, we will evaluate the performance of our merging algorithm under latency-sensitive traffic conditions representative of 5G URLLC applications. The goal is to assess how our merging mechanism interacts with cooperative scheduling protocols and whether it can maintain SLA guarantees while supporting real-time mobile traffic over PON.

\section{Conclusion}
% In this work, we presented a real time heuristic stateful algorithm for upstream scheduling in a TWDM multi-tenant PON, capable of satisfying SLAs. We show the role that ONU tuning time plays in terms of latency performance in the trade-off between higher capacity single channel and lower capacity multi-channel systems. Our current results show that tuning times between 250 ns and 1 $\mu s$ would be required to maintain satisfactory performance in multi-tenant, multi-channel PON systems. We show that the performance of the algorithm in terms of SLA compliancy is highly affected by the distribution patterns of the upstream network traffic from the ONUs. We show that the runtime of the algorithm is directly proportional with the available network bandwidth. Finally, we conclude from our study that our merging algorithm is able to run in 5G and future 6G low-latency services and applications with minimal additional latency.
In this work, we presented a real-time heuristic stateful algorithm for upstream scheduling in a TWDM multi-tenant PON, capable of satisfying SLAs. We show the role that ONU tuning time plays in terms of latency performance in the trade-off between higher capacity single-channel and lower capacity multi-channel systems. Our current results show that tuning times between 250~ns and 1~$\mu$s would be required to maintain satisfactory performance in multi-tenant, multi-channel PON systems.

We show that the performance of the algorithm in terms of SLA compliancy is highly affected by the distribution patterns of the upstream network traffic from the ONUs. We show that the runtime of the algorithm is directly proportional with the available network bandwidth. 

Our findings also highlight that, although the DTWA algorithm performs well under typical traffic conditions, its SLA compliance drops under bursty or self-similar traffic patterns. We proposed a classification-driven approach to mitigate this variance by identifying bursty ONUs and assigning them transmission windows more appropriately. This not only helps maintain fairness but also reduces resource contention and improves compliance in highly dynamic scenarios.

While the DTWA algorithm demonstrates sub-10~$\mu$s runtime performance in software simulation for rates up to 50\,Gbps, our runtime scaling analysis shows that higher data rate systems will likely require hardware acceleration. Future integration with PON testbeds based on Intel’s DPDK is planned to evaluate the algorithm under continuous real-time operation and hardware-imposed constraints.

Additionally, we plan to evaluate the feasibility of deploying our algorithm in O-RAN fronthaul use cases. Given that URLLC applications have tight latency budgets, we aim to validate whether our merging mechanism can meet end-to-end QoS constraints when deployed within a cooperative DBA framework, as discussed in Section~\ref{introduction}~\nameref{introduction} and Section~\ref{experiment}~\nameref{experiment}.

Finally, we conclude from our study that our merging algorithm is able to run in 5G and future 6G low-latency services and applications with minimal additional latency.

\section*{Acknowledgments} 
This publication has emanated from research conducted with the financial support of Taighde Éireann – Research Ireland under Grant numbers 12/RC/2276\_p2 (IPIC), 18/RI/5721 (OpenIreland) and  13/RC/2077\_p2 (CONNECT).

% Bibliography
\bibliography{sample}

\end{document}